\documentclass[useAMS,usenatbib]{mnras}
\usepackage{natbib}
\usepackage{amsmath}
\usepackage{url}
\usepackage{longtable}
\usepackage{aas_macros}
\usepackage{amssymb}
\usepackage{graphicx}
\usepackage{deluxetable}
\usepackage{pdflscape}
\usepackage{multirow}
\usepackage{color}

\newcommand{\err}[2]{\ensuremath{^{+#1}_{-#2}}}

\def\lsim{\hbox{\rlap{\raise 0.425ex\hbox{$<$}}\lower 0.65ex\hbox{$\sim$}}}
\def\gsim{\hbox{\rlap{\raise 0.425ex\hbox{$>$}}\lower 0.65ex\hbox{$\sim$}}}

\newcommand{\lvcmone}{\ensuremath{2.02\err{0.58}{0.34}}}
\newcommand{\lvcmtwo}{\ensuremath{1.35\err{0.26}{0.27}}}
\newcommand{\lvcchirp}{\ensuremath{1.437\err{0.022}{0.020}}}
\newcommand{\lvcq}{\ensuremath{0.67\err{0.29}{0.25}}}
\newcommand{\lvcmtot}{\ensuremath{3.39\err{0.32}{0.11}}}
\newcommand{\lvcchi}{\ensuremath{0.06\err{0.11}{0.05}}}
\newcommand{\lvcdl}{\ensuremath{159\err{69}{71}}}

\newcommand{\bnsmone}{\ensuremath{1.85\err{0.27}{0.19}}}
\newcommand{\bnsmtwo}{\ensuremath{1.47\err{0.16}{0.18}}}
\newcommand{\bnschirp}{\ensuremath{1.436\err{0.022}{0.019}}}
\newcommand{\bnsq}{\ensuremath{0.79\err{0.18}{0.19}}}
\newcommand{\bnsmtot}{\ensuremath{3.33\err{0.10}{0.06}}}
\newcommand{\bnschi}{\ensuremath{0.03\err{0.04}{0.03}}}
\newcommand{\bnsdl}{\ensuremath{162\err{67}{73}}}

\newcommand{\bnsmchmone}{\ensuremath{2.03\err{0.15}{0.14}}}
\newcommand{\bnsmchmtwo}{\ensuremath{1.35 \pm 0.09}}
\newcommand{\bnsmchchirp}{\ensuremath{1.433\err{0.023}{0.019}}}
\newcommand{\bnsmchq}{\ensuremath{0.67\err{0.10}{0.08}}}
\newcommand{\bnsmchmtot}{\ensuremath{3.38\err{0.08}{0.07}}}
\newcommand{\bnsmchchi}{\ensuremath{0.05 \pm 0.03}}
\newcommand{\bnsmchdl}{\ensuremath{171\err{66}{76}}}

\newcommand{\bnsqmone}{\ensuremath{1.70\err{0.17}{0.06}}}
\newcommand{\bnsqmtwo}{\ensuremath{1.60\err{0.05}{0.14}}}
\newcommand{\bnsqchirp}{\ensuremath{1.436\err{0.021}{0.019}}}
\newcommand{\bnsqq}{\ensuremath{0.94\err{0.05}{0.16}}}
\newcommand{\bnsqmtot}{\ensuremath{3.31 \pm 0.05}}
\newcommand{\bnsqchi}{\ensuremath{0.02\err{0.03}{0.02}}}
\newcommand{\bnsqdl}{\ensuremath{161\err{66}{70}}}

\newcommand{\nsbhmone}{\ensuremath{2.19\err{0.21}{0.17}}}
\newcommand{\nsbhmtwo}{\ensuremath{1.26\err{0.10}{0.08}}}
\newcommand{\nsbhchirp}{\ensuremath{1.438\err{0.021}{0.020}}}
\newcommand{\nsbhq}{\ensuremath{0.57\err{0.10}{0.08}}}
\newcommand{\nsbhmtot}{\ensuremath{3.46\err{0.13}{0.09}}}
\newcommand{\nsbhchi}{\ensuremath{0.07\err{0.04}{0.03}}}
\newcommand{\nsbhdl}{\ensuremath{154\err{70}{69}}}

\title[Updated Parameters and the EM Counterpart of GW190425]{Updated
  Parameter Estimates for GW190425 Using Astrophysical Arguments and
  Implications for the Electromagnetic Counterpart}

\def\ucsc{1}
\def\carn{2}
\def\nbi{3}

\begin{document}

\author[Foley et~al.]{Ryan~J.~Foley$^{\ucsc}$\thanks{E-mail:foley@ucsc.edu},
David~A.~Coulter$^{\ucsc}$,
Charles~D.~Kilpatrick$^{\ucsc}$,
Anthony~L.~Piro$^{\carn}$,
\newauthor
Enrico~Ramirez-Ruiz$^{\ucsc,\nbi}$,
Josiah~Schwab$^{\ucsc}$\\
$^{\ucsc}$Department of Astronomy and Astrophysics, University of California, Santa Cruz, CA 95064, USA\\
$^{\carn}$The Observatories of the Carnegie Institution for Science, 813 Santa Barbara Street, Pasadena, CA 91101, USA\\
$^{\nbi}$Niels Bohr Institute, University of Copenhagen, Blegdamsvej 17, DK-2100 Copenhagen, Denmark
}

\date{Accepted  . Received   ; in original form  }
\pagerange{\pageref{firstpage}--\pageref{lastpage}} \pubyear{2016}
\maketitle
\label{firstpage}

\begin{abstract}
  The progenitor system of the compact binary merger GW190425 had a
  total mass of $3.4\err{0.3}{0.1}$~M$_{\sun}$ (90th-percentile
  confidence region, with individual component masses of $m_{1} =
  \lvcmone$ and $m_{2} = \lvcmtwo$~M$_{\sun}$) as measured from its
  gravitational wave signal.  This mass is significantly different
  from the Milky Way (MW) population of binary neutron stars (BNSs)
  that are expected to merge in a Hubble time and from that of the
  first BNS merger, GW170817.  Here we explore the expected
  electromagnetic signatures of such a system.  We make several
  astrophysically motivated assumptions to further constrain the
  parameters of GW190425.  By simply assuming that both components
  were NSs, we reduce the possible component masses significantly,
  finding $m_{1} = \bnsmone$~M$_{\sun}$ and $m_{2} =
  \bnsmtwo$~M$_{\sun}$.  However if the GW190425 progenitor system was
  a NS-black hole merger, we find best-fitting parameters $m_{1} =
  \nsbhmone$~M$_{\sun}$ and $m_{2} = \nsbhmtwo$~M$_{\sun}$.  For a
  well-motivated BNS system where the lighter NS has a mass similar to
  the mass of non-recycled NSs in MW BNS systems, we find $m_{1} =
  \bnsmchmone$~M$_{\sun}$ and $m_{2} = \bnsmchmtwo$~M$_{\sun}$,
  corresponding to only 7\% mass uncertainties and reducing the
  90th-percentile mass range to 32\% and 34\% the size of the original
  range, respectively.  For all scenarios, we expect a prompt collapse
  of the resulting remnant to a black hole.  Examining detailed models
  with component masses similar to our best-fitting results, we find
  the electromagnetic counterpart to GW190425 is expected to be
  significantly redder and fainter than that of GW170817.  We find
  that almost all reported observations used to search for an
  electromagnetic counterpart for GW190425 were too shallow to detect
  the expected counterpart.  If the LIGO-Virgo Collaboration promptly
  provides the chirp mass, the astronomical community can adapt their
  observations to improve the likelihood of detecting a counterpart
  for similarly ``high-mass'' BNS systems.
\end{abstract}



\defcitealias{Abbott20}{LVC20}

\section{Introduction}\label{s:intro}

The detection of gravitational waves (GWs) from the merger of compact
objects has unveiled new populations of black holes
\citep[BHs;][]{Abbott16:gw, Abbott19:sample} and now neutron stars
(NSs).  Observations of GW190425, the second probable BNS merger
detected, by the LIGO-Virgo Collaboration \citep[LVC;][hereafter
\citetalias{Abbott20}]{Abbott20} indicate that it was likely a binary
NS (BNS) merger, but with the total mass of the system,
$3.4\err{0.3}{0.1}$~M$_{\sun}$, being significantly larger than any of
the known BNS systems in the Milky Way \citep[MW; e.g.,][]{Ozel16,
  Farrow19}.  This puzzling system has already led to questions about
its formation scenario \citep{Romero-Shaw20, Safarzadeh20}.

GW190425 is especially intriguing since the first BNS merger,
GW170817, had a total mass consistent with the MW BNS population
\citep{Abbott17:gw170817}.  The relative rates of GW170817-like and
GW190425-like events are comparable \citepalias{Abbott20}, indicating
that significant selection effects, dramatically different delay
times, and/or significant environmental differences are necessary to
produce such mergers at similar rates yet have no ``high-mass''
systems in the MW sample.

Ideally, we would have detected an electromagnetic (EM) counterpart to
GW190425 to pinpoint its location and glean additional information
about the system.  For GW170817, \citet{Coulter17} discovered the
optical counterpart, AT~2017gfo (also known as Swope Supernova Survey
2017a or SSS17a), giving way to numerous studies in the areas of
high-energy astrophysics \citep{Goldstein17,
  Murguia-Berthier17:sss17a}, nuclear physics \citep{Margalit17,
  Abbott18:eos, Annala18, Radice18:eos}, general relativity
\citep{Barack19}, $r$-process nucleosynthesis \citep{Kasen17,
  Kilpatrick17:gw, Rosswog18}, compact-object formation channels
\citep{Blanchard17, Levan17, Pan17:gw, Ramirez-Ruiz19}, and cosmology
\citep{Abbott17:h0, Guidorzi17}.

However, no such counterpart has been confidently discovered for
GW190425 \citep{Antier19, Coughlin19, Lundquist19, Hosseinzadeh19},
perhaps partly because of the large localization area for GW190425
\citepalias[90th-percentile region of 8,284~deg$^{2}$;][]{Abbott20},
high distance (\lvcdl~Mpc), and that a significant fraction of the
localization region was either Sun constrained or at low Galactic
latitudes.  An alternative possibility is that the counterpart was
intrinsically faint, as indicated by the luminosity function of
kilonovae associated with short gamma-ray bursts \citep{Gompertz18,
  Ascenzi19}.

\citetalias{Abbott20} performed a sophisticated and detailed analysis
of the GW strain data and provided posterior distributions for
parameters that describe the
system\footnote{https://dcc.ligo.org/LIGO-P2000026/public}.  Their
approach was to make minimal assumptions and allow the data constrain
the parameter estimates.  However, part of the resulting allowed
parameter space is inconsistent with separate astrophysical
constraints on NSs and BHs.  Here, we apply additional astrophysical
constraints to significantly reduce the allowed parameter space for
GW190425.  With the more constraining data, we examine the
characteristics of a possible EM counterpart, finding that it likely
was significantly fainter and redder than AT~2017gfo.

We introduce our astrophysically motivated assumptions and apply them
to the GW190425 data in Section~\ref{s:sys}.  In Section~\ref{s:imp},
we explore how the updated parameter estimates affect the properties
of an EM counterpart.  In Section~\ref{s:disc}, we discuss additional
implications of the GW190425 system.  We conclude in
Section~\ref{s:conc}.


\section{Physically Motivated Systems}\label{s:sys}

In this section, we make several arguments to reduce the parameter
space of possible progenitor systems for GW190425.  We first separate
different scenarios into BNS and NSBH mergers.

In the former scenario, high spins and a particularly high mass for
$m_{1}$ are ruled out by physical arguments.  We are further able to
restrict the parameter space by examining two possible progenitor
systems, an equal-mass system and a system including a NS whose mass
is consistent with the non-recycled NSs in MW BNS systems (a mass
close to a Chandrasekhar mass).

Similar physical arguments require that the NSBH scenario have an
extreme mass ratio.  Again, assuming that the lighter object is
consistent with the MW BNS population provides stringent constraints.

\subsection{Is GW190425 a BNS or NSBH Merger?}\label{ss:fate}

Given the total mass of the system, GW190425 is likely a BNS merger,
but an NSBH merger cannot be ruled out by the GW data alone \citep[see
also][]{Han20}.  Unless there is a population of very low-mass BHs
(i.e., $M < 1.7$~M$_{\sun}$), GW190425 was not a BBH merger, and we do
not further consider that scenario.

We distinguish if the primary object is a BH based on its mass.  We
assume that if $m_{1} \le M_{\rm TOV}$, the Tolman-Oppenheimer-Volkoff
(TOV) mass \citep{Oppenheimer39, Tolman39}, then the primary is a NS.
We assume it is a BH if $m_{1} > M_{\rm TOV}$.  Because we do not
precisely know the equation of state for nuclear matter, we must
constrain this maximum mass through observations.  We use two pulsar
mass measurements for this constraint. Of well-measured pulsars, the
most important for this quantity is the millisecond pulsar J0740+6620,
which has the highest measured mass, $M_{0740} =
2.14\err{0.10}{0.09}$~M$_{\sun}$ \citep{Cromartie20}.  We also include
J0348+0432 with a mass of $M_{0348} = 2.01 \pm 0.04$~M$_{\sun}$, which
because of its small mass uncertainty is particularly important for
excluding the low-mass tail.  Critically for the NSBH case, we assume
$M_{\rm TOV} \ge M_{0704}$ and $M_{0348}$.

We also assume $M_{\rm TOV}$ is less than the maximum value determined
from requiring a hypermassive NS after the merger of GW170817.
Several studies have performed similar analyses \citep{Margalit17,
  Shibata17, Shibata19, Ruiz18, Ai19}, all finding $M_{\rm TOV}
\lesssim 2.1$ -- 2.3~$M_{\sun}$ (at 90th-percentile confidence).  We
use the distribution of limits from these sources to produce a
one-sided Gaussian distribution to predict the upper limit for $M_{\rm
  TOV}$, finding it is best fit with a central value of 2.10 and a
width of 0.06~M$_{\sun}$, corresponding to a 90th-percentile limit of
2.23~M$_{\sun}$.  Notably this distribution is also consistent with
the measured mass of the pulsar J0740+6620.

Additionally, we assume that the less massive component of the
GW190425 system is a NS, requiring $m_{2} \ge M_{\rm min,~NS}$, the
minimum mass of a NS.  For this minimum mass, we adopt the mass of the
unseen companion in J0453+1559, $M = 1.174$~M$_{\sun}$
\citep{Martinez15}.  \citet{Suwa18} suggest a stellar evolution
pathway for producing such a low-mass NS, although see
\citet{Tauris19} for an alternative scenario in which this object is
instead a white dwarf.

Starting with the PhenomPv2NRT \citep{Dietrich19} high-spin posterior
distributions \citepalias[corresponding to their preferred
model]{Abbott20}, we resample the posterior distributions with
additional conditions.  Whenever we assume a component is a NS, we
require that its mass be $M_{\rm min,~NS} \le M \le M_{\rm TOV,~max}$.
When we assume a component is a black hole, we assume $M > M_{\rm
  TOV,~min}$.

We additionally apply a prior on the spin for any assumed NS
component.  Gravitational wave data does not directly constrain the
spin, but rather
\begin{equation}
  \mathbf{\chi} = \frac{c S}{G m^{2}},
\end{equation}
where $S$ is the amplitude of the spin vector and $m$ is the NS mass.
This results in a degeneracy between NS spin and mass.  Equivalently,
one can write $\chi$ as
\begin{equation}
  \chi = \frac{I \Omega c}{G m^{2}},
\end{equation}
where $I$ is the moment of inertia and $\Omega$ is the spin frequency.
For a NS with a millisecond period and reasonable NS equations of
state, we find a maximum $\chi$ of $\sim$0.2, significantly less than
the ``high-spin'' prior of $\chi < 0.89$ used by \citetalias{Abbott20}.
To be conservative, we apply a prior of $\chi < 0.4$ for any NS
component.

\begin{figure}
\begin{center}
\includegraphics[angle=0,width=3.2in]{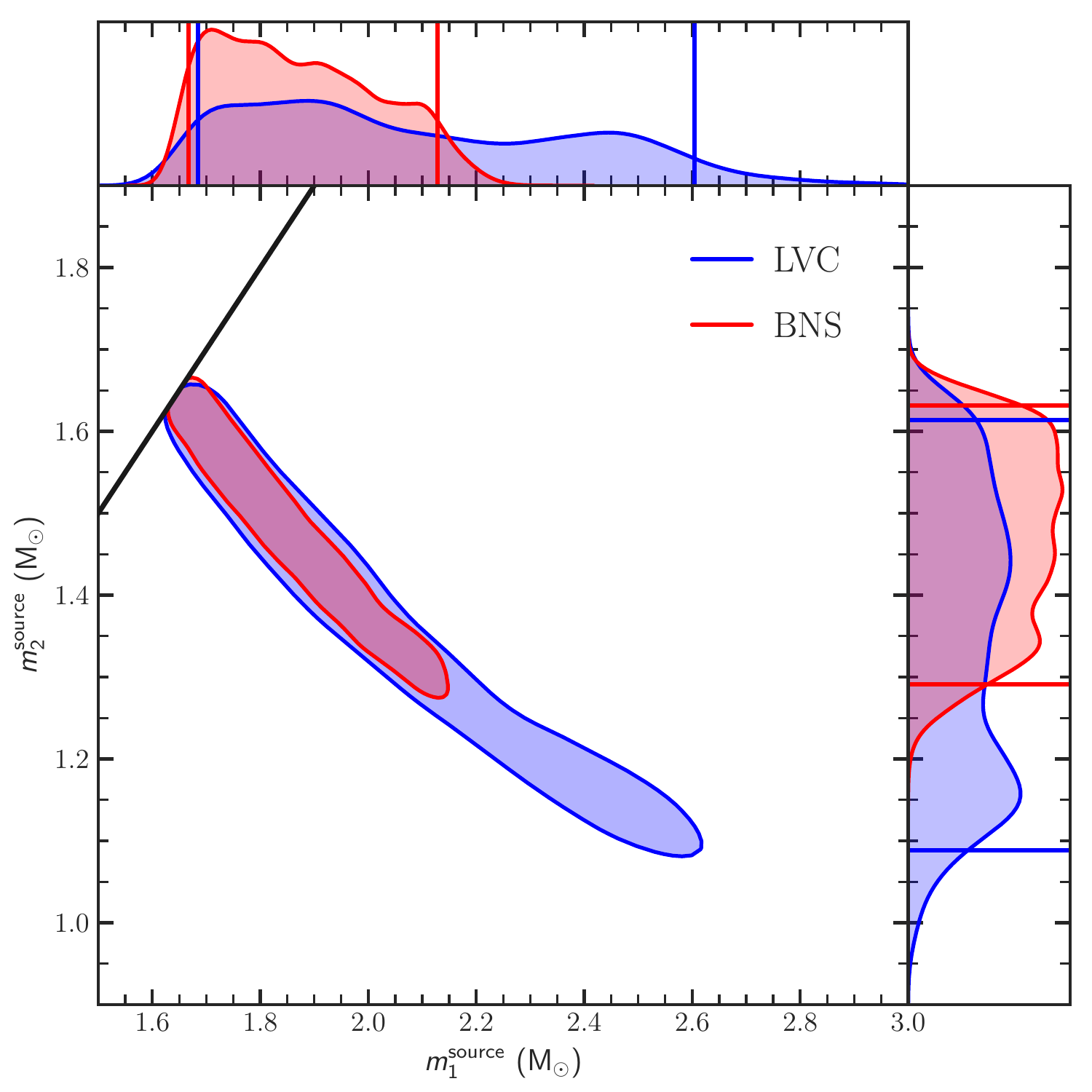}
\caption{Posterior distributions of the component masses $m_{1}$ and
  $m_{2}$ in the source frame.  The blue contours represent the
  \citetalias{Abbott20} high-spin prior distributions, while the red
  contours are recovered by assuming GW190425 was a BNS merger.  The
  solid black line represents systems with equal mass.  Vertical lines
  in the one-dimensional plots enclose 90\% of the probability with
  the distributions scaled to have the same total
  area.}\label{f:bns_mass}
\end{center}
\end{figure}

Without any assumptions about the nature of the progenitor system,
\citetalias{Abbott20} found a total system mass of \lvcmtot~M$_{\sun}$
and component masses of $m_{1} = \lvcmone$~M$_{\sun}$ and $m_{2} =
\lvcmtwo$~M$_{\sun}$ (all 90th-percentile confidence regions and
measured in the source frame).  If we assume that GW190425 was a BNS
system, the allowed parameter space reduces to $M_{\rm tot} =
\bnsmtot$~M$_{\sun}$ and component masses of $m_{1} =
\bnsmone$~M$_{\sun}$ and $m_{2} = \bnsmtwo$~M$_{\sun}$, making the
total mass uncertainty only 37\% that of its original measurement.  We
present the updated mass constraints in Figure~\ref{f:bns_mass} and
Table~\ref{t:prop}.

As noted by \citetalias{Abbott20}, $m_{1}$ and $\chi$ are highly
correlated for GW190425.  Not including the $\chi$ constraint results
in a $\chi$ distribution for the more massive component that contains
very little probability at $>$0.4.  However, $\sim$5\% of the $\chi$
distribution for the less-massive component is at $>$0.4.  We
therefore include this condition despite its limited predictive power.

For the NSBH system, we can apply a separate constraint that there is
no tidal deformation for the more massive component (i.e.,
$\Lambda_{1} = 0$).  We performed the analysis both with and without
this constraint, finding no significant difference.  We therefore do
not include the constraint.

\begin{figure}
\begin{center}
\includegraphics[angle=0,width=3.2in]{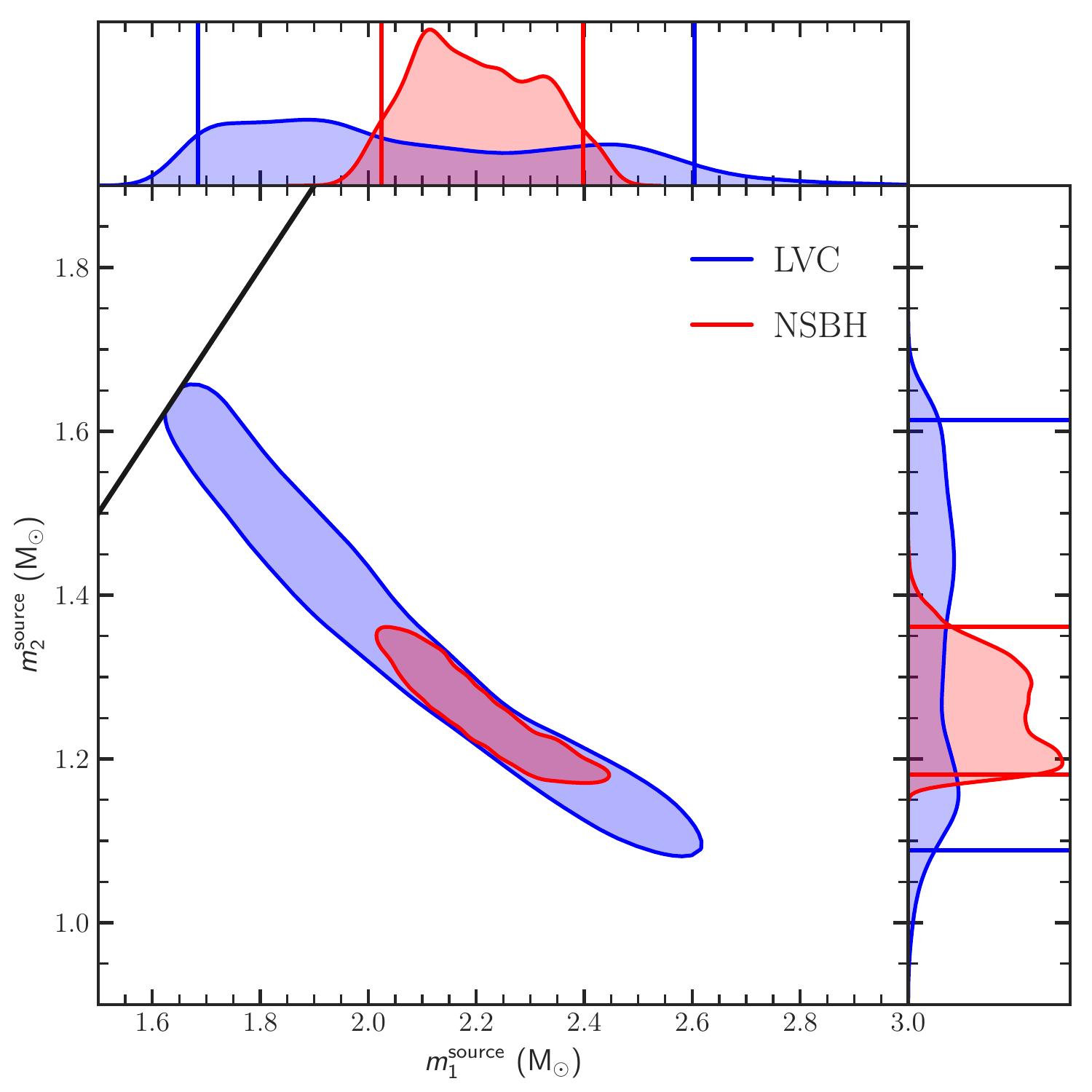}
\caption{Same as Figure~\ref{f:bns_mass}, but where the solid red
  contours are recovered by assuming GW190425 was an NSBH
  merger.}\label{f:nsbh_mass}
\end{center}
\end{figure}

If we assume that GW190425 was a NSBH merger, we find $M_{\rm tot} =
\nsbhmtot$~M$_{\sun}$ and component masses of $m_{1} =
\nsbhmone$~M$_{\sun}$ and $m_{2} = \nsbhmtwo$~M$_{\sun}$, reducing the
total mass uncertainty to 56\% that of its original uncertainty.  We
present the updated mass constraints in Figure~\ref{f:nsbh_mass}.

\subsection{A Chandrasekhar-mass Neutron Star}

The individual NSs in Milky Way BNS systems have a mass distribution
centered at 1.33~M$_{\sun}$ and a standard deviation of
0.09~M$_{\sun}$ \citep{Ozel16}, indicating that such NSs have a
preference for masses similar to the Chandrasekhar mass\footnote{We
  remind the reader that while the Chandrasekhar mass might be related
  to the birth of a NS, it does not physically constrain its mass.
  However, we use the name as a reference to the approximate mass of
  the population of MW BNSs.}.  Separately, core-collapse modeling
indicates that many remnants should be close to a Chandrasekhar mass
\citep{Sukhbold16}.  This is especially expected in the case of an
electron-capture supernova \citep{Kitaura06} or the accretion-induced
collapse of a WD \citep{Timmes96}.  There is a correlation between NS
mass and kick velocity \citep{Podsiadlowski04}, providing an
additional reason why at least the second NS in a merging BNS system
would have a mass on the low-end of viable NS masses.

Given the total mass of the GW190425 system, having one
Chandrasekhar-mass NS would imply that the second NS would have $m
\approx 2$~M$_{\sun}$.  Such a massive NS is obviously not observed in
Milky Way BNS systems, however, NSs that massive are known, including
in NS-WD systems \citep{Antoniadis16}.

Combined, a 1.4+2.0~M$_{\sun}$ system is a natural choice for the
GW190425 progenitor system.  Such masses are consistent with either a
BNS or NSBH system and is close to the best-fitting masses from GW
data alone (1.35+2.02~M$_{\sun}$).  We note, however, that such a
system is somewhat in tension with the low-spin prior ($\chi < 0.05$)
assumed by \citetalias{Abbott20}.

\begin{figure}
\begin{center}
\includegraphics[angle=0,width=3.2in]{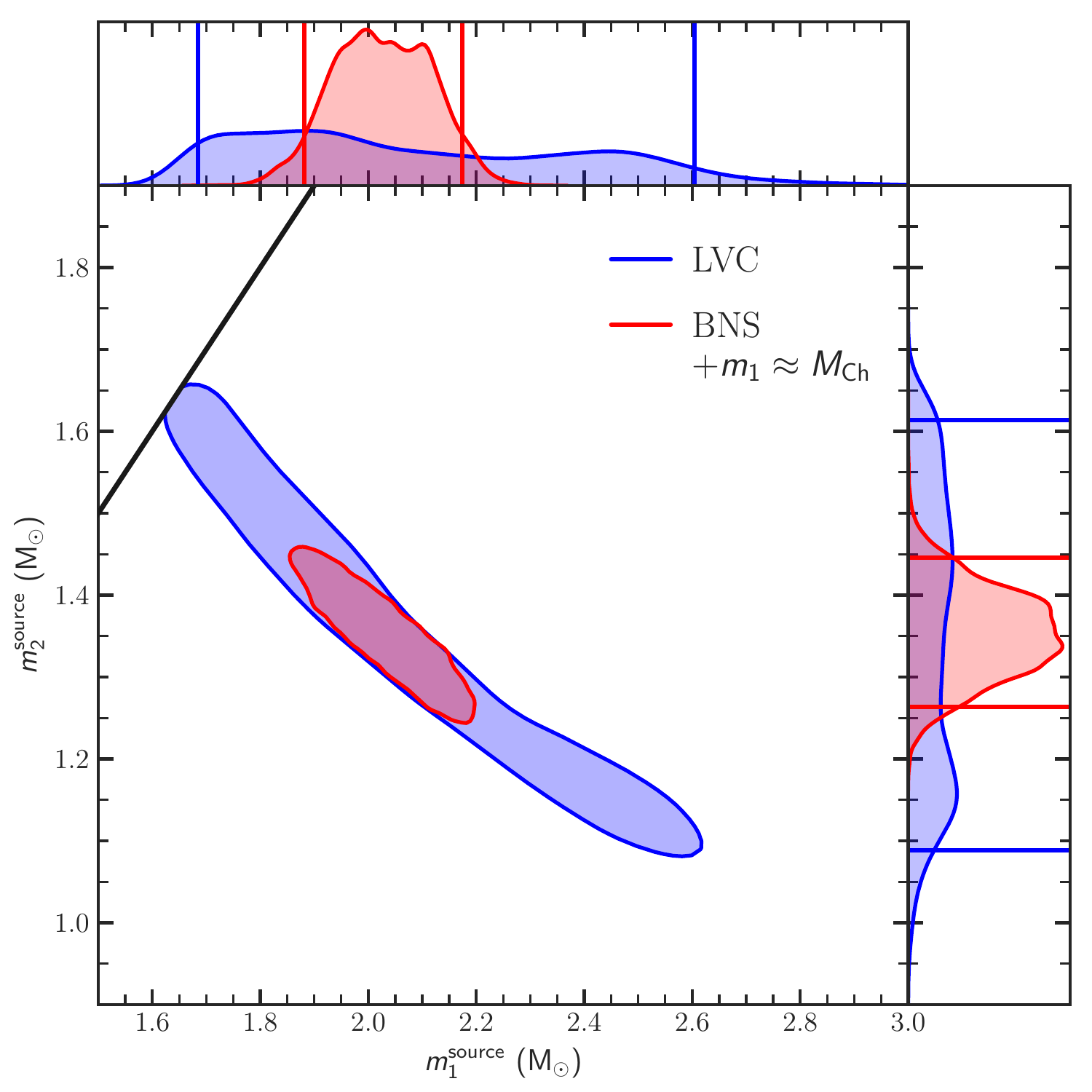}
\caption{Same as Figure~\ref{f:bns_mass}, but where the solid red
  contours are recovered by assuming GW190425 was a BNS merger where
  the lighter object was NS drawn from the \citet{Farrow19}
  non-recycled NS mass distribution.}\label{f:bnsmch_mass}
\end{center}
\end{figure}

Motivated by these considerations, we use the best-fitting population
to the non-recycled pulsars in MW BNS systems as determined by
\citet{Farrow19}.  They find that a uniform prior with a lower bound
of $1.16\err{0.01}{0.02}$~M$_{\sun}$ and an upper bound of
$1.42\err{0.04}{0.02}$~M$_{\sun}$ is a better fit than a Gaussian
distribution for the non-recycled NSs.  Using this prior and assuming
a BNS system, we find $m_{1} = \bnsmchmone$~M$_{\sun}$ and $m_{2} =
\bnsmchmtwo$~M$_{\sun}$.  While these values are consistent with the
\citetalias{Abbott20} values and from assuming a BNS system, the mass
ranges are significantly reduced.  We present the updated mass
constraints in Figure~\ref{f:bnsmch_mass}.

In this scenario, we find a mass ratio of $q = \bnsmchq$, indicating a
system far from equal mass, but intriguingly close to the best-fitting
value without any astrophysical assumptions ($q = \lvcq$.
Additionally, we find that the luminosity distance is somewhat, but
insignificantly, larger with these assumption ($D_{L} = \bnsmchdl$~Mpc
with and $D_{L} = \lvcdl$~Mpc without).

For the NSBH scenario, the posterior distributions do not change
significantly when including this additional assumption.  As can be
seen from Figure~\ref{f:nsbh_mass}, the secondary mass is already
constrained to within the \citet{Farrow19} mass distribution with the
basic NSBH assumptions.

\subsection{An Equal-mass System}

All known MW BNS systems have mass ratios of $q \approx 1$, with the
lowest measured value for a system expected to merge in a Hubble time
of $0.75 \pm 0.05$ \citep{Ferdman18} and the second-lowest being $0.92
\pm 0.01$ \citep{Ferdman14}.  While none have a total mass of
$>$2.9~M$_{\sun}$, there is a clear preference for equal-mass systems
in this population.  Additionally, the GW170817 BNS system and nearly
all BBH systems are also consistent with their components having equal
mass \citep{Abbott19:params, Abbott19:sample}.  It is clear that
Nature produces many compact binary systems with near-equal mass.

We use the distribution of mass ratios from the MW BNS systems
as determined by \citet{Farrow19} to investigate the implications for
GW190425.  The 90th-percentile prior lower bound on $q$ is 0.836.
This configuration requires a low spin to be consistent with the GW
data.

\begin{figure}
\begin{center}
\includegraphics[angle=0,width=3.2in]{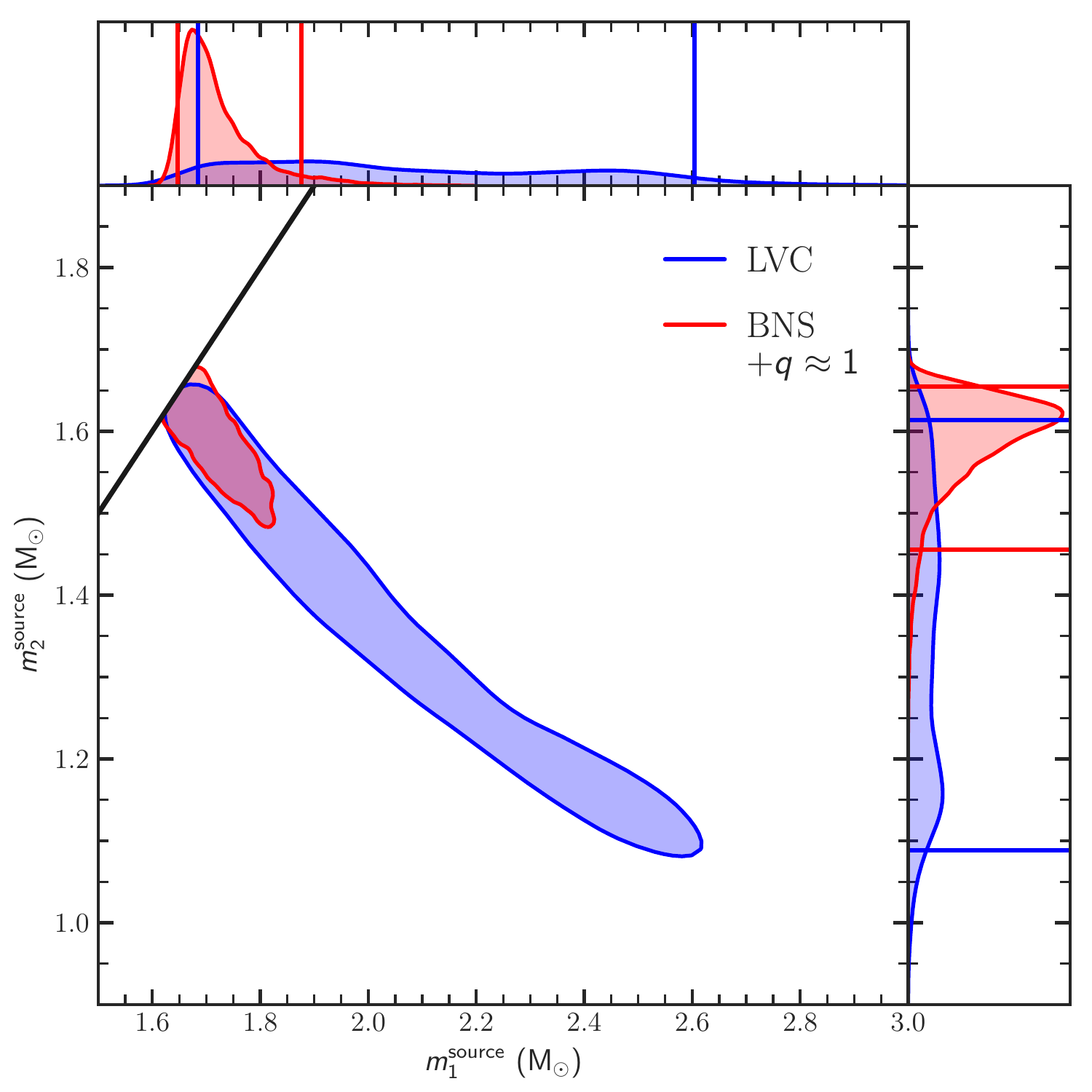}
\caption{Same as Figure~\ref{f:bns_mass}, but where the solid red
  contours are recovered by assuming GW190425 was a near-equal-mass
  BNS merger.}\label{f:q_mass}
\end{center}
\end{figure}

Under the assumption that GW190425 was as BNS with roughly equal-mass
components, the parameter space is especially constrained.  We find
the components had masses of $m_{1} = \bnsqmone $~M$_{\sun}$ and
$m_{2} = \bnsqmtwo$~M$_{\sun}$.  We present the updated mass
constraints in Figure~\ref{f:q_mass}.  The mass ratio is constrained
to $q = \bnsqq$ and $\chi_{\rm eff} = \bnsqchi$, consistent with
expectations and the low-spin prior from \citetalias{Abbott20}.

\section{Astrophysical Implications}\label{s:imp}

In this section, we detail the implications of the tighter priors
assumed in Section~\ref{s:sys}.

\subsection{Remnant}\label{ss:remnant}

The source-frame chirp mass for GW190425 of $\mathcal{M} =
\lvcchirp$~M$_{\sun}$ is sufficiently large that if GW190425 came from
a BNS system, it likely resulted in a prompt collapse to a black hole
\citep[see e.g.,][]{Piro17}.  We can assess this by examining the
remnant mass compared to the TOV mass and the radius of a
1.6~M$_{\sun}$ NS, $R_{1.6}$.  Following \citet{Bauswein13} and
\citet{Bauswein17}, we expect prompt collapse if
\begin{equation}
  M_{\rm tot} > M_{\rm thresh} \approx \left ( 2.38 - 3.606 \frac{G
      M_{\rm TOV}}{c^{2} R_{1.6}} \right ) M_{\rm TOV}.
\end{equation}
The remnant should have underwent prompt collapse if $M_{\rm tot}$ is
larger than the maximum value of $M_{\rm thresh}$ given current
constraints on $M_{\rm TOV}$ and $R_{1.6}$.  For $M_{\rm TOV} =
2.17$~M$_{\sun}$ \citep{Margalit17} and $R_{1.6} = 13.5$~km
\citep{De18}, we find $M_{\rm thresh,~max} = 3.3$~M$_{\sun}$.

The remnant for GW190425 likely had a mass similar to the threshold
mass.  From GW data alone, GW190425 had $M_{\rm tot} =
\lvcmtot$~M$_{\sun}$.  However, we should examine only BNS mergers
since NSBH systems will never have a NS remnant, even if short lived.
In Section~\ref{ss:fate}, we found that if GW190425 is a BNS system,
it had $M_{\rm tot} = \bnsmtot$~M$_{\sun}$.  Assuming that the lighter
component had a mass similar to the Galactic mass distribution, we
find GW190425 had $M_{\rm tot} = \bnsmchmtot$~M$_{\sun}$.  As noted
below, it is unlikely that it ejected more than 0.04~M$_{\sun}$.
Additionally, $M_{\rm thresh}$ may be as low as $\sim$2.7~M$_{\sun}$
given current constraints on $M_{\rm TOV}$ and $R_{1.6}$.  We
therefore believe it is likely that the remnant of GW190425 promptly
collapsed to a black hole.

This basic analysis is consistent with that of \citetalias{Abbott20},
which found a 97\% chance of a prompt collapse using recent
simulations, a range of equations of state, and their high-spin
priors.  A prompt collapse implies that GW190425 was likely
accompanied by a short gamma-ray burst \citep{Murguia-Berthier14},
which was not confidently seen (although, see \citealt{Pozanenko19});
however, the data are not constraining \citep{Song19}.

For the remainder of this work, we assume that if GW190425 was a BNS
merger, it underwent a prompt collapse.  Even if it produced a
hypermassive NS, it would have been sufficiently massive to collapse
quickly, making the resulting ejecta qualitatively similar to the
prompt-collapse scenario.

\subsection{Ejecta}

\citet{Rosswog13} produced several dynamical merger models of compact
binary mergers \citep[see also][]{Korobkin12}.  The models include
both BNS and NSBH mergers over a large mass range.  We use these
models to estimate the amount of dynamical ejecta and its velocity.
In particular, the 1.4+2.0 and 1.6+1.6 BNS merger models produced
0.039 and 0.020~M$_{\sun}$ of ejecta with an average velocity of 0.15
and 0.11$c$, respectively.  There are no low-mass NSBH models, but the
1.4+5.0 NSBH (corresponding to the least-massive NSBH merger
presented) model produced 0.024~M$_{\sun}$ with an average velocity of
0.15$c$, and lower-mass models are expected to produce less ejecta
with a smaller average velocity.

Other studies have examined a range of NS masses (and equations of
state), but do not fully cover the range of possible systems for
GW190425 \citep[e.g.,][]{Bauswein13, Radice17}, and we therefore focus
on the \citet{Rosswog13} results for consistency.  We note, however,
that these alternative simulations generally find ejecta masses that
are smaller by a factor of $\gtrsim$2 depending primarily on the
assumed NS equation of state.  Because of these differences, we
consider the \citet{Rosswog13} models as having optimistic ejecta
masses and GW190425 may have ejected significantly less material than
corresponding models.

We argue in Section~\ref{ss:remnant} that GW190425 likely underwent a
prompt collapse to a BH.  In this scenario, all ejecta (regardless of
exact origin) is expected to have low electron fraction, $Y_{e}$.

\subsection{Electromagnetic Counterparts}

AT~2017gfo, the optical counterpart to GW170817 \citep{Coulter17}, was
relatively bright at all optical wavelengths, peaking at $M \approx
-16$~mag \citep{Andreoni17, Cowperthwaite17:gw, Drout17, Evans17,
  Kasliwal17, McCully17, Smartt17, Tanvir17, Utsumi17, Valenti17}.  It
faded extremely quickly in blue bands but was longer lived in the NIR
\citep[e.g.,][]{Siebert17}.  This broad-band behavior requires ejecta
with a distribution of lanthanide fractions, $X_{\rm lan}$ that is
often parameterized as separate components each with a single
lanthanide fraction \citep[e.g.,][]{Drout17, Villar17}.

The lanthanide fraction is directly related to the electron fraction,
$Y_{e}$, of the material, with $Y_{e} \lesssim 0.25$ material
producing a significant fraction of lanthanides \citep{Lippuner15}.
Ejecta created through separate processes during the merger are
expected to have different $Y_{e}$ \citep{Metzger14} with the tidal
debris having $Y_{e} \lesssim 0.1$ \citep[e.g.,][]{Rosswog05} and a
neutrino-irradiated disk wind having $Y_{e} \gtrsim 0.25$
\citep{Fernandez13}.  The amount of ejecta from the disk wind and its
$Y_{e}$ depends critically on the mass of the remnant and lifetime of
a possible hypermassive or supramassive NS \citep{Metzger14}.  The
bright blue colors of AT~2017gfo require a short-lived hypermassive NS
\citep{Margalit17, Murguia-Berthier17:sss17a}.

As discussed in Section~\ref{ss:remnant}, GW190425 likely underwent a
prompt collapse and had low $Y_{e}$, and thus the ejecta should have a
high $X_{\rm lan}$.  All models considered exclusively produce
low-$Y_{e}$ ejecta that correspond to high lanthanide fractions of
$\log X_{\rm lan} \approx -2$.

\citet{Kasen17} produced light curves for kilonovae with a range of
$M_{\rm ej}$, $v$, and $X_{\rm lan}$.  \citet{Kilpatrick17:gw} used a
combination of two of these models corresponding to low-lanthanide and
high-lanthanide components to match the light curves of AT~2017gfo.
This is confirmation that kilonova observations can be reproduced by
theoretical models.

\begin{figure}
\begin{center}
\includegraphics[angle=0,width=3.2in]{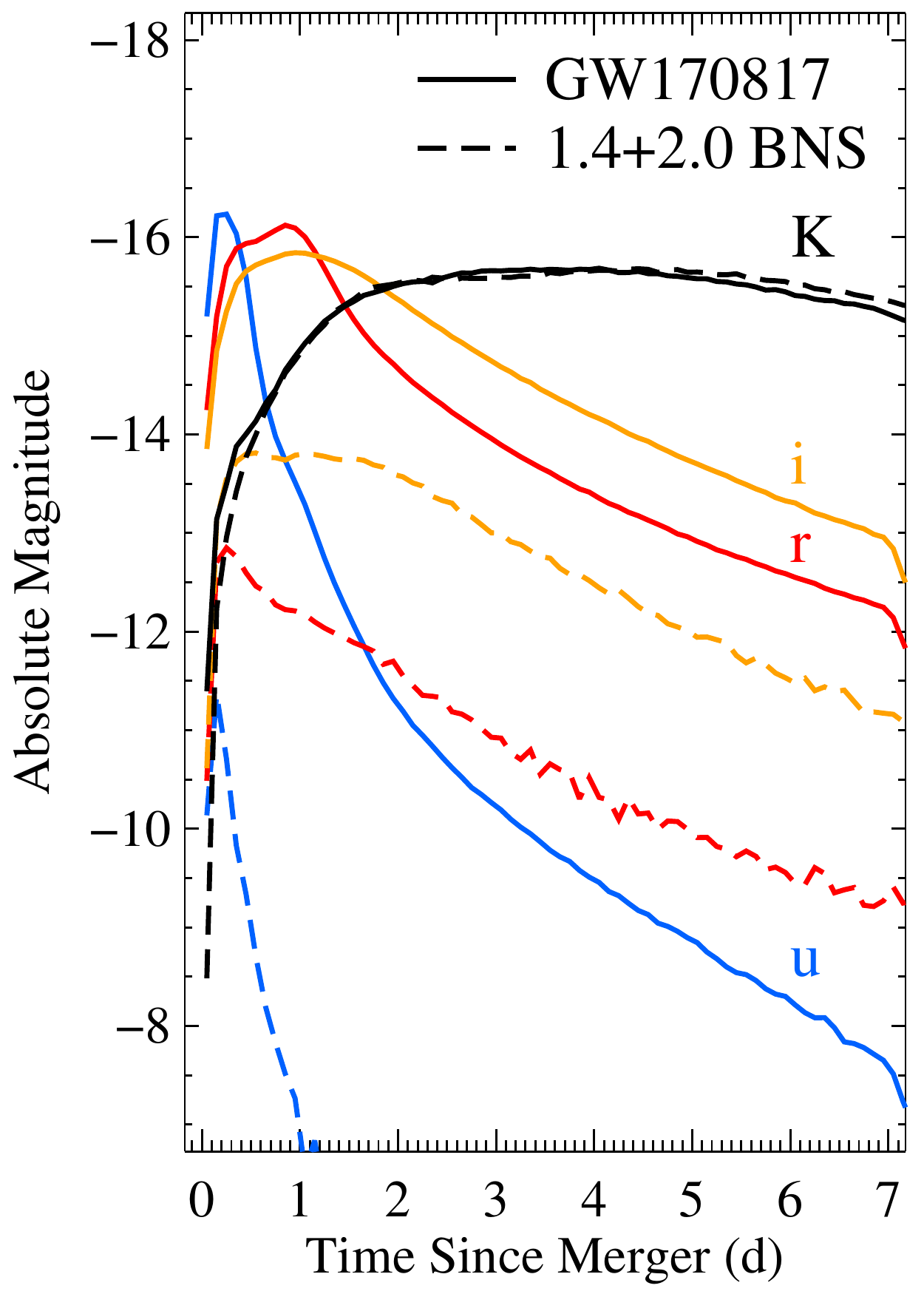}
\caption{$uriK$ (blue, red, orange, and black curves) absolute light
  curves for kilonova models \citep{Kasen17}.  The solid curve is a
  model matched to the observations of GW170817
  \citep{Kilpatrick17:gw} and contains both blue, low lanthanide
  fraction and red, high lanthanide fraction components.  The dashed
  curves represent a model with $M_{\rm ej} = 0.04$~M$_{\sun}$, $v =
  0.15 c$, and $\log X_{\rm lan} = -2$, similar parameters to
  hydrodynamical models of a 1.4+2.0~M$_{\sun}$ BNS merger
  \citep{Rosswog13}, a likely scenario for GW190425.}\label{f:kn}
\end{center}
\end{figure}

For comparison, we examine light curves for a kilonova with $M_{\rm
  ej} = 0.025$, 0.03, and 0.04~M$_{\sun}$; $v = 0.1$, 0.15, and
0.15$c$; and $\log X_{\rm lan} = -2$ (for all models), respectively.
These models are similar to the \citet{Rosswog13} models for 1.6+1.6,
1.2+2.0, and 1.4+2.0~M$_{\sun}$ BNS mergers.  The first two models are
also somewhat similar to the 1.4+5.0~M$_{\sun}$ NSBH merger.  We
display the light curves corresponding to the 1.4+2.0~M$_{\sun}$the
model and the \citet{Kilpatrick17:gw} model for AT~2017gfo in
Figure~\ref{f:kn}.  While the NIR light curves of these two models are
nearly identical, the optical light curves are significantly different
with the 1.4+2.0~M$_{\sun}$ BNS merger producing a kilonova with
significantly less optical luminosity.  The other models, which have
less ejecta mass, produce significantly less luminous kilonovae, with
the 1.6+1.6~M$_{\sun}$ model being $\sim$1~mag fainter in $i$ at peak
than the 1.4+2.0~M$_{\sun}$ model.  The 1.4+2.0~M$_{\sun}$ model is
the most luminous of models consistent with GW190425.

\begin{figure}
\begin{center}
\includegraphics[angle=0,width=3.2in]{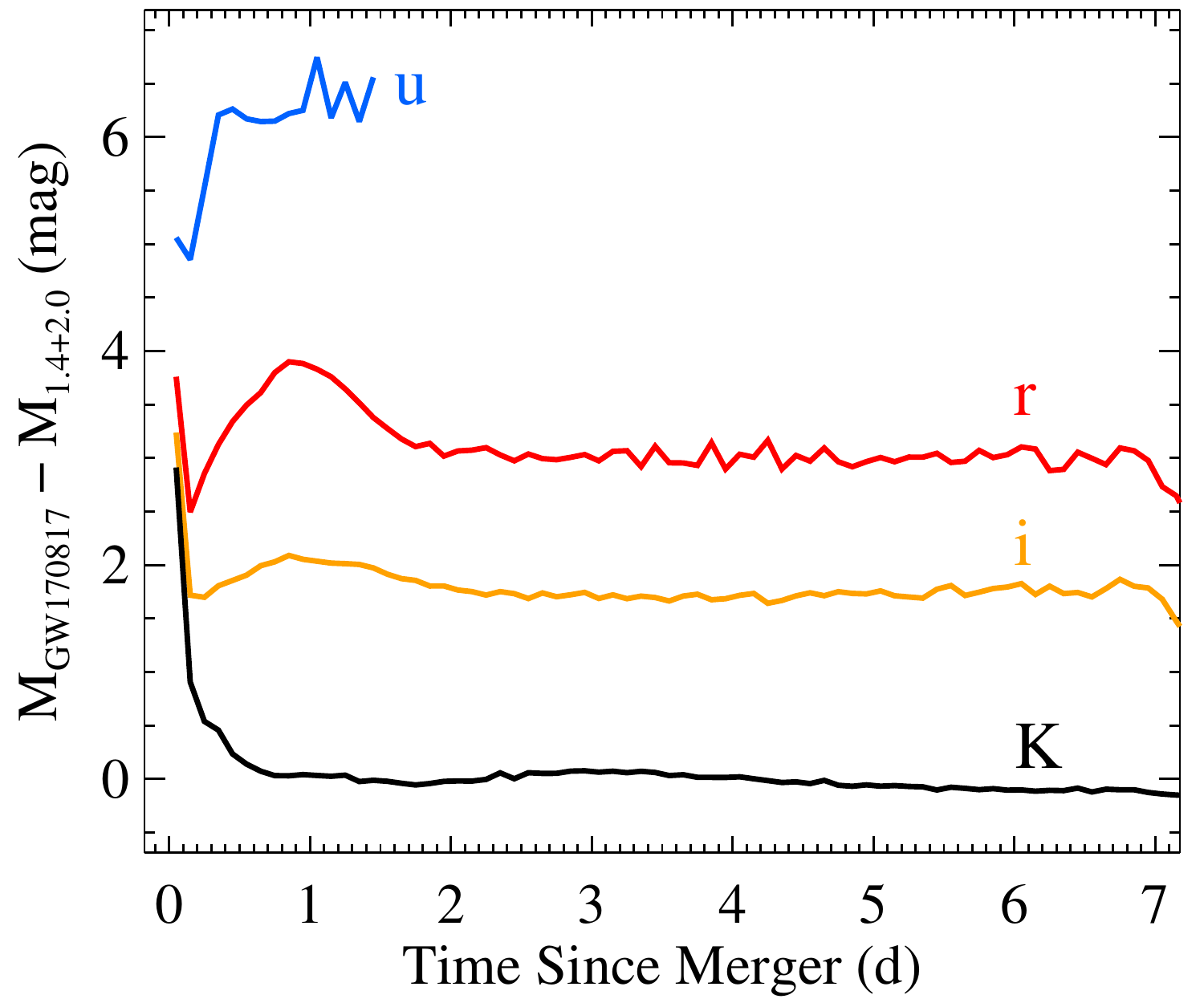}
\caption{Difference in absolute magnitude between kilonova models
  \citep{Kasen17} for GW170817 \citep{Kilpatrick17:gw} and a model
  with $M_{\rm ej} = 0.04$~M$_{\sun}$, $v = 0.15 c$, and $\log X_{\rm
    lan} = -2$, similar parameters to hydrodynamical models of a
  1.4+2.0~M$_{\sun}$ BNS merger \citep{Rosswog13}, a likely scenario
  for GW190425.  The blue, red, orange, and black curves correspond to
  the $uriK$ bands, respectively.  The $u$ band is shown only for the
  first 1.5~d after merger, after which time it is so faint in the
  1.4+2.0~M$_{\sun}$ model that there are insufficient photon
  statistics to create a reasonable light curve.  In the optical
  bands, we expect GW190425 to be several magnitudes less luminous
  than GW170817.}\label{f:kn_diff}
\end{center}
\end{figure}

In Figure~\ref{f:kn_diff}, we display the differences between the two
models.  As seen in Figure~\ref{f:kn}, the $K$-band light curves are
nearly identical.  However, the 1.4+2.0~M$_{\sun}$ BNS merger kilonova
is $\sim$6, 3, and 2~mag less luminous than GW170817 in $uri$ over the
first week after the merger.

\section{Discussion}\label{s:disc}

\subsection{Parameters}

The astrophysically motivated assumptions reduced the uncertainties on
several parameters.  As these assumptions were never in tension with
the observations, the parameters did not shift dramatically from the
\citetalias{Abbott20} results.

While our assumptions significantly affected the mass-related
parameters (each component mass, total mass, and mass ratio),
moderately affected $\chi_{\rm eff}$, and had a minor effect on the
distance, it did not change most parameters such as the tidal
deformation or inclination.  As such, our study cannot improve our
understanding of the nuclear equation of state or more precisely
constrain its location.

Nevertheless, the mass and spin parameters are important for
understanding the origin of the system and its ultimate fate.  The
BNS, BNS with $m_{2} \approx M_{\rm Ch}$, and NSBH scenarios are all
consistent with $m_{1} = 2.0$~M$_{\sun}$, $m_{2} = 1.3$~M$_{\sun}$,
and $\chi_{\rm eff} = 0.05$.  These values are consistent with a
separate analysis specifically examining the NSBH case \citep{Han20},
although the values presented here are more precise.  The $q \approx
1$ scenario disfavors such a system and instead prefers $m_{1} \approx
m_{2} \approx 1.65$~M$_{\sun}$ and $\chi_{\rm eff} = 0.02$.

For all scenarios except the equal-mass case, we find a relatively
high $\chi_{\rm eff}$ that is inconsistent with MW BNS systems.  Such
a constrained system (masses known to $\sim$10\% and spin constraints)
should guide explorations of possible pathways to creating such a
system.  Similarly, there must only be a few possible ways to create a
low-spin, high-but-equal-mass system.

\subsection{Searching for an EM Counterpart to GW190425}

In the minutes after the merger of GW190425, the LVC publicly
announced its detection as a likely BNS merger\footnote{This is not an
  assessment of the structure of the individual components.  The
  classification scheme used considers any system where each component
  has $m < 3$~M$_{\sun}$ a BNS.} with a low false-alarm rate \citep[1
per 69,834 years;][]{LVC19:0425_1}.  They quickly released a
three-dimensional probability map of the location of the GW emission
with a two-dimensional 90th-percentile localization region of
10,183~deg$^{2}$ which was revised about a day later to having a size
of 7,461~deg$^{2}$ \citep{LVC19:0425_2}; the final map released in
2020 January has a 8,284~deg$^{2}$ two-dimensional 90th-percentile
localization region \citepalias{Abbott20}.

Although this region covers $\sim$1/5 of the entire sky, several
groups searched for an EM counterpart \citep[e.g.,][]{Antier19,
  Coughlin19, Hosseinzadeh19, Lundquist19}.  Collectively, $\sim$50\%
of the two-dimensional localization area was observed in the days
following the merger.

We highlight three sets of observations: \citet{Coughlin19} observed
8000~deg$^{2}$ in $gr$ with a limiting magnitude of $\sim$21 and
2400~deg$^{2}$ in $J$ with a limiting magnitude of $\sim$15.5;
\citet{Hosseinzadeh19} observed 67 galaxies in $i$ with a limiting
magnitude of $\sim$22.5; and \citet{Tohuvavohu19} observed 408
galaxies in $u$ with a limiting magnitude of 21.1.  These observations
correspond to the widest/reddest, deepest, and bluest searches,
respectively.

GW190425 has a distance of \lvcdl~Mpc, corresponding to a
distance modulus of $36.01\err{0.78}{1.28}$~mag.  Assuming no dust
extinction (in the MW, host galaxy, or circumstellar environment), the
observations listed above have absolute magnitude limits of about
$-15.0$~mag in wide-field $gr$, $-20.5$~mag in wide-field $J$,
$-13.5$~mag in the deepest $i$, and $-14.9$~mag in $u$, with an
uncertainty of $\sim$1~mag depending on the distance to GW190425.
Taking the distance into account, we find that the \citet{Coughlin19}
$g$-, $r$-, and $J$-band observations could detect a GW190425-like
1.4+2.0~M$_{\sun}$ BNS merger to $D = 46$, 59, and 15~Mpc,
respectively, indicating that those observations were insensitive to
this event.  The \citet{Tohuvavohu19} $u$-band observations could
detect a GW190425-like 1.4+2.0~M$_{\sun}$ BNS merger to $D = 31$~Mpc,
again indicating that they were not sensitive enough to detect the EM
counterpart of GW190425.  The \citet{Hosseinzadeh19} $i$-band
observations were sufficiently deep to detect GW190425 to
$\sim$150~Mpc, not quite the median distance, if observed at the time
of peak luminosity.

\subsection{Future Searches}

The community's search for an EM counterpart to GW190425 was likely
inefficient.  The difference in efficiency between the GW190425 search
and the extremely successful search for the counterpart of GW170817 is
primarily caused by the physical differences of the systems and the
likely prompt collapse of the GW190425 system to a black hole.
(Differences in success also likely depended on the localization
area.) This resulted in an absolute magnitude difference in the $r$
band (the most commonly used filter for searching) of $\Delta M_{r} =
3$~mag.  The difference in distance between the two events results in
another 1.7--3.8~mag difference.  Combined, the EM counterpart of
GW190425 was likely 5--7~mag fainter than AT~2017gfo in the $r$ band.

While AT~2017gfo had similar peak luminosities in all optical bands,
the EM counterpart to GW190425 was likely much redder having $r-i
\approx 1$~mag at peak and $r-i \approx 2$~mag at +2~d after merger
($g-i \approx 1.5$ and 5~mag at comparable times).  It is therefore
likely more efficient to search in redder bands for similar events,
even at the loss of detector sensitivity.

\subsection{Chemical Enrichment}

A BNS or NSBH merger is expected to disperse heavy, $r$-process
elements into its environments \citep{Lattimer74, Freiburghaus99}.
The rate of enrichment from these events is important for
understanding galaxy evolution and if additional sources of
$r$-process material are necessary \citep[e.g.,][]{Shen15,
  VandeVoort15, VandeVoort19, Naiman18}.

AT~2017gfo is often modeled as having two components, each with
different lanthanide fractions.  For instance, \citet{Kilpatrick17:gw}
modeled AT~2017gfo as having a ``red'' component with $M_{\rm ej} =
0.035$~M$_{\sun}$, $v = 0.15 c$, and $\log X_{\rm lan} = -2$ and a
``blue'' component with $M_{\rm ej} = 0.025$~M$_{\sun}$, $v = 0.25 c$,
and $\log X_{\rm lan}$ ranging from $-4$ to $-6$.  From this, they
find a total amount of $r$-process material ejected of $M_{\rm r-p}
\approx 0.06$~M$_{\sun}$.  This value is consistent with other
estimates \citep{Drout17, Kasliwal17, Smartt17, Tanaka17, Tanvir17,
  Villar17}.

We estimate GW190425 ejected $0.03 \pm 0.01$~M$_{\sun}$ of material
with $\log X_{\rm lan} \approx -2$, similar to the ``red'' component
of GW170817/AT~2017gfo, but substantially less than the total mass
ejected.  This estimate is larger than the stringent minimum mass
required to effectively pollute low metallicity, $r$-process enhanced
stars \citep{Macias18}.

\citetalias{Abbott20} determined the individual merger rates for
``GW170817-like'' and ``GW190425-like'' events as well as a combined
event rate assuming both are drawn from the same population.
Following the procedure of \citet{Kilpatrick17:gw}, but using the
individual rates and total ejecta masses for each event, we find a
total of $5\err{11}{4} \times 10^{4}$~M$_{\sun}$ of $r$-process
material produced in a MW-like galaxy over $10^{10}$~yr.  This amount
is comparable to that estimated from the Solar $r$-process abundance
and total mass of stars and gas in the MW \citep{Grevesse07, Kafle14},
indicating that compact-object mergers alone are sufficient for
producing the majority of $r$-process material in the Universe.

While GW170817 likely produced both ``light'' ($A < 140$) and
``heavy'' ($A > 140$) $r$-process material, GW190425 likely produced
primarily heavy $r$-process material.  Using a similar argument as
above, but using the GW170817-like rate for the blue component of
GW170817 and the combined rate for the red component of GW170817 and
GW190425, we find compact-object mergers have produced $4\err{12}{2}
\times 10^{4}$~M$_{\sun}$ of light and $0.4\err{1.1}{0.2} \times
10^{4}$~M$_{\sun}$ of heavy $r$-process material produced throughout
the history of the MW.  Roughly 10\% of $r$-process material produced
in NS mergers, averaged over all events, is $A > 140$ nuclei.

Mergers that eject primarily dynamical material with low $Y_{e}$ are
expected to be an excellent site for actinide production, perhaps
being a way to produce the abundances seen in ``actinide-boost'' stars
\citep[e.g.,][]{Hill02}.  Notably, whatever process enriches
actinide-boost stars, it must also produce a similar abundance pattern
to the process that enriches normal low-metallicity stars for lighter
elements \citep{Roederer09}.  While GW190425-like events may be the
source of heavy elements in actinide-boost stars, the dynamical ejecta
of NSs is expected to {\it overproduce} actinides \citep[although see
\citealt{Wanajo14} for a nuanced examination of the $Y_{e}$
distribution in the dynamical ejecta]{Holmbeck19}.  This may indicate
additional components to the GW190425 ejecta or GW190425-like events
being rarer in the MW than the measured rate would imply.

\subsection{Expanding the Compact-Object Parameter Space}

The key observation for GW190425 is the high total mass.  As we have
described in detail, the system could range from an equal-mass BNS
system with component masses of $\sim$1.7~M$_{\sun}$ each to a NSBH
with $m_{1} = 2.4$~M$_{\sun}$ and $q = 0.5$.

The NSBH scenario requires a BH significantly less massive than the
population of known MW BHs \citep{Ozel10, Farr11} and would exist in
the ``mass gap.''  However, the mass is so low that it is unlikely to
be the result of a previous NS merger.  It therefore appears that such
a scenario requires producing such a low-mass BH from core collapse.
A population of $\sim$2~M$_{\sun}$ BHs are not currently predicted
from most core-collapse simulations \citep[e.g.,][]{Ugliano12},
however stellar evolution modeling indicates a possible pathway
\citep{Ertl19}.  Further examination of theoretical models to
understand the likelihood of this possibility is required for this
scenario to be truly viable.

Additionally, there are only a few simulations of high total mass or
extreme mass ratio compact object mergers.  Additional simulations of
BNS mergers from 1.6+1.6~M$_{\sun}$ to 1.3+2.1~M$_{\sun}$ and NSBH
mergers with 1.4+2.0~M$_{\sun}$ to 1.2+2.4~M$_{\sun}$ are necessary to
fully understand the properties of the GW190425 ejecta.  Future GW
events may have slightly higher total mass indicating even higher mass
ratios.  Extending current merger simulations to higher total masses
and more extreme mass ratios will be important for precise modeling of
this new population of binary systems.


\section{Conclusions}\label{s:conc}

The discovery of GW190425, the second likely BNS merger, was a
landmark event, unveiling a new population of binary compact object
systems not yet discovered in the MW.  We can infer from the exquisite
GW data that the total system mass is significantly different from the
known population of MW BNSs.

Extending that analysis, we use astrophysical knowledge to further
constrain the parameters of GW190425.  Depending on the exact
assumptions, we can reduce the parameter space to $<$30\% that of
\citetalias{Abbott20}.  We detail three relevant scenarios that can
then be used in future studies to understand the formation of the
progenitor system, focus merger models, and enhance population
studies.

If GW190425 was a BNS system where the lighter component has a mass
similar to the non-recycled NSs in MW BNS systems, we find $m_{1} =
\bnsmchmone$, $m_{2} = \bnsmchmtwo$, and $m_{\rm tot} =
\bnsmchmtot$~M$_{\sun}$.  Such a system would have a small mass ratio
($q = \bnsmchq$) and a relatively high $\chi_{\rm eff}$ of \bnsmchchi\
(all uncertainties being 90th percentile).

However, if we assume that GW190425 had a mass ratio similar to that
of MW BNS systems that merge within a Hubble time, we find $m_{1} =
\bnsqmone$, $m_{2} = \bnsqmtwo$, and $m_{\rm tot} =
\bnsqmtot$~M$_{\sun}$.  This highly constrained system is also
expected to have a small $\chi_{\rm eff}$ of $\bnsqchi$.

Finally, if GW190425 was a NSBH system, we find $m_{1} = \nsbhmone$,
$m_{2} = \nsbhmtwo$, and $m_{\rm tot} = \nsbhmtot$~M$_{\sun}$.  This
system requires an extreme mass ratio of $q = \nsbhq$ and a relatively
high $\chi_{\rm eff}$ of \nsbhq.

Regardless of the exact scenario, we expect the remnant to either
already be a BH or promptly collapse to a BH.  Again, regardless of
the system, we expect $\sim$0.03~M$_{\sun}$ of lanthanide-rich
ejecta.  Using updated event rate estimates and these ejecta
characteristics, we find that compact-object mergers produce a
significant fraction --- and perhaps essentially all --- $r$-process
material in the Universe.

The expected kilonova associated with GW190425 is expected to have
similar NIR properties as AT~2017gfo, but should be significantly
fainter in the optical.  As a result, we believe essentially no
follow-up observation obtained to find the electromagnetic counterpart
was constraining.

Optical/infrared searches could choose appropriate filters and adjust
exposure times to optimally search for GW170817-like and GW190425-like
events {\it if} LVC promptly released the chirp mass.  This
information is available during the initial analysis and would
significantly improve chances of detecting counterparts while
improving the efficiency of observatories.

\section*{Acknowledgements}

\bigskip

This work was only possible because of the LVC and its members.  Their
tireless work operating the GW detectors, analyzing the data, and
providing data to the community has sped discovery in the nascent
field.  We especially thank them for providing the posterior data sets
soon after \citetalias{Abbott20} was made public.

Much of this manuscript was conceived while R.J.F.\ attended the
Hirschegg 2020 workshop.  We thank the organizers and attendees,
especially B.\ C\^ot\'e, S.\ De, R.\ Essick, R.\ Gamba, G.\
Raaijmakers, S.\ Reddy, and L.\ Roberts for enlightening conversations
about GW190425.  We also thank P.\ Brady, D.\ Lin, P.\ Macias, B.\
Margalit, M.\ Safarzadeh, and S.\ Woosley for additional help and
conversations.

The UCSC team is supported in part by NASA grant NNG17PX03C, NSF grant
AST-1815935, the Gordon \& Betty Moore Foundation, the Heising-Simons
Foundation, and by fellowships from the Alfred P.\ Sloan Foundation
and the David and Lucile Packard Foundation to R.J.F.  E.R.-R.\ thanks
the Heising-Simons Foundation, the Danish National Research Foundation
(DNRF132) and NSF (AST-1911206 and AST-1852393) for support.  D.A.C.\
acknowledges support from the National Science Foundation Graduate
Research Fellowship under Grant DGE1339067.  J.S.\ is also supported
by the A.\ F.\ Morrison Fellowship at Lick Observatory.


\bibliographystyle{mnras}
\bibliography{../astro_refs}

\begin{thebibliography}{}
\makeatletter
\relax
\def\mn@urlcharsother{\let\do\@makeother \do\$\do\&\do\#\do\^\do\_\do\%\do\~}
\def\mn@doi{\begingroup\mn@urlcharsother \@ifnextchar [ {\mn@doi@}
  {\mn@doi@[]}}
\def\mn@doi@[#1]#2{\def\@tempa{#1}\ifx\@tempa\@empty \href
  {http://dx.doi.org/#2} {doi:#2}\else \href {http://dx.doi.org/#2} {#1}\fi
  \endgroup}
\def\mn@eprint#1#2{\mn@eprint@#1:#2::\@nil}
\def\mn@eprint@arXiv#1{\href {http://arxiv.org/abs/#1} {{\tt arXiv:#1}}}
\def\mn@eprint@dblp#1{\href {http://dblp.uni-trier.de/rec/bibtex/#1.xml}
  {dblp:#1}}
\def\mn@eprint@#1:#2:#3:#4\@nil{\def\@tempa {#1}\def\@tempb {#2}\def\@tempc
  {#3}\ifx \@tempc \@empty \let \@tempc \@tempb \let \@tempb \@tempa \fi \ifx
  \@tempb \@empty \def\@tempb {arXiv}\fi \@ifundefined
  {mn@eprint@\@tempb}{\@tempb:\@tempc}{\expandafter \expandafter \csname
  mn@eprint@\@tempb\endcsname \expandafter{\@tempc}}}

\bibitem[\protect\citeauthoryear{{Abbott} et~al.,}{{Abbott}
  et~al.}{2016}]{Abbott16:gw}
{Abbott} B.~P.,  et~al., 2016, \mn@doi [Physical Review Letters]
  {10.1103/PhysRevLett.116.061102}, \href
  {http://adsabs.harvard.edu/abs/2016PhRvL.116f1102A} {116, 061102}

\bibitem[\protect\citeauthoryear{{Abbott} et~al.,}{{Abbott}
  et~al.}{2017a}]{Abbott17:gw170817}
{Abbott} B.~P.,  et~al., 2017a, \mn@doi [Physical Review Letters]
  {10.1103/PhysRevLett.119.161101}, \href
  {http://adsabs.harvard.edu/abs/2017PhRvL.119p1101A} {119, 161101}

\bibitem[\protect\citeauthoryear{{Abbott} et~al.,}{{Abbott}
  et~al.}{2017b}]{Abbott17:h0}
{Abbott} B.~P.,  et~al., 2017b, \mn@doi [\nat] {10.1038/nature24471}, \href
  {http://adsabs.harvard.edu/abs/2017Natur.551...85A} {551, 85}

\bibitem[\protect\citeauthoryear{{Abbott} et~al.,}{{Abbott}
  et~al.}{2018}]{Abbott18:eos}
{Abbott} B.~P.,  et~al., 2018, \mn@doi [\prl] {10.1103/PhysRevLett.121.161101},
  \href {https://ui.adsabs.harvard.edu/abs/2018PhRvL.121p1101A} {121, 161101}

\bibitem[\protect\citeauthoryear{{Abbott} et~al.,}{{Abbott}
  et~al.}{2019}]{Abbott19:params}
{Abbott} B.~P.,  et~al., 2019, \mn@doi [Physical Review X]
  {10.1103/PhysRevX.9.011001}, \href
  {https://ui.adsabs.harvard.edu/abs/2019PhRvX...9a1001A} {9, 011001}

\bibitem[\protect\citeauthoryear{{Ai}, {Gao}  \& {Zhang}}{{Ai}
  et~al.}{2019}]{Ai19}
{Ai} S.,  {Gao} H.,   {Zhang} B.,  2019, arXiv e-prints, \href
  {https://ui.adsabs.harvard.edu/abs/2019arXiv191206369A} {p. arXiv:1912.06369}

\bibitem[\protect\citeauthoryear{{Andreoni} et~al.,}{{Andreoni}
  et~al.}{2017}]{Andreoni17}
{Andreoni} I.,  et~al., 2017, \mn@doi [\pasa] {10.1017/pasa.2017.65}, \href
  {http://adsabs.harvard.edu/abs/2017PASA...34...69A} {34, e069}

\bibitem[\protect\citeauthoryear{{Annala}, {Gorda}, {Kurkela}  \&
  {Vuorinen}}{{Annala} et~al.}{2018}]{Annala18}
{Annala} E.,  {Gorda} T.,  {Kurkela} A.,   {Vuorinen} A.,  2018, \mn@doi [\prl]
  {10.1103/PhysRevLett.120.172703}, \href
  {https://ui.adsabs.harvard.edu/abs/2018PhRvL.120q2703A} {120, 172703}

\bibitem[\protect\citeauthoryear{{Antier} et~al.,}{{Antier}
  et~al.}{2019}]{Antier19}
{Antier} S.,  et~al., 2019, \mn@doi [\mnras] {10.1093/mnras/stz3142}, \href
  {https://ui.adsabs.harvard.edu/abs/2019MNRAS.tmp.2740A} {p.~2740}

\bibitem[\protect\citeauthoryear{{Antoniadis}, {Tauris}, {Ozel}, {Barr},
  {Champion}  \& {Freire}}{{Antoniadis} et~al.}{2016}]{Antoniadis16}
{Antoniadis} J.,  {Tauris} T.~M.,  {Ozel} F.,  {Barr} E.,  {Champion} D.~J.,
  {Freire} P. C.~C.,  2016, arXiv e-prints, \href
  {https://ui.adsabs.harvard.edu/abs/2016arXiv160501665A} {p. arXiv:1605.01665}

\bibitem[\protect\citeauthoryear{{Ascenzi} et~al.,}{{Ascenzi}
  et~al.}{2019}]{Ascenzi19}
{Ascenzi} S.,  et~al., 2019, \mn@doi [\mnras] {10.1093/mnras/stz891}, \href
  {http://adsabs.harvard.edu/abs/2019MNRAS.tmp..865A} {}

\bibitem[\protect\citeauthoryear{{Barack} et~al.,}{{Barack}
  et~al.}{2019}]{Barack19}
{Barack} L.,  et~al., 2019, \mn@doi [Classical and Quantum Gravity]
  {10.1088/1361-6382/ab0587}, \href
  {https://ui.adsabs.harvard.edu/abs/2019CQGra..36n3001B} {36, 143001}

\bibitem[\protect\citeauthoryear{{Bauswein}, {Baumgarte}  \&
  {Janka}}{{Bauswein} et~al.}{2013}]{Bauswein13}
{Bauswein} A.,  {Baumgarte} T.~W.,   {Janka} H.-T.,  2013, \mn@doi [Physical
  Review Letters] {10.1103/PhysRevLett.111.131101}, \href
  {http://adsabs.harvard.edu/abs/2013PhRvL.111m1101B} {111, 131101}

\bibitem[\protect\citeauthoryear{{Bauswein}, {Just}, {Janka}  \&
  {Stergioulas}}{{Bauswein} et~al.}{2017}]{Bauswein17}
{Bauswein} A.,  {Just} O.,  {Janka} H.-T.,   {Stergioulas} N.,  2017, \mn@doi
  [\apjl] {10.3847/2041-8213/aa9994}, \href
  {http://adsabs.harvard.edu/abs/2017ApJ...850L..34B} {850, L34}

\bibitem[\protect\citeauthoryear{{Blanchard} et~al.,}{{Blanchard}
  et~al.}{2017}]{Blanchard17}
{Blanchard} P.~K.,  et~al., 2017, \mn@doi [\apjl] {10.3847/2041-8213/aa9055},
  \href {http://adsabs.harvard.edu/abs/2017ApJ...848L..22B} {848, L22}

\bibitem[\protect\citeauthoryear{{Coughlin}, {Dietrich}, {Heinzel}, {Khetan},
  {Antier}, {Christensen}, {Coulter}  \& {Foley}}{{Coughlin}
  et~al.}{2019}]{Coughlin19}
{Coughlin} M.~W.,  {Dietrich} T.,  {Heinzel} J.,  {Khetan} N.,  {Antier} S.,
  {Christensen} N.,  {Coulter} D.~A.,   {Foley} R.~J.,  2019, arXiv e-prints,
  \href {https://ui.adsabs.harvard.edu/abs/2019arXiv190800889C} {}

\bibitem[\protect\citeauthoryear{{Coulter} et~al.,}{{Coulter}
  et~al.}{2017}]{Coulter17}
{Coulter} D.~A.,  et~al., 2017, \mn@doi [Science] {10.1126/science.aap9811},
  \href {http://adsabs.harvard.edu/abs/2017Sci...358.1556C} {358, 1556}

\bibitem[\protect\citeauthoryear{{Cowperthwaite} et~al.,}{{Cowperthwaite}
  et~al.}{2017}]{Cowperthwaite17:gw}
{Cowperthwaite} P.~S.,  et~al., 2017, \mn@doi [\apjl]
  {10.3847/2041-8213/aa8fc7}, \href
  {http://adsabs.harvard.edu/abs/2017ApJ...848L..17C} {848, L17}

\bibitem[\protect\citeauthoryear{{Cromartie} et~al.,}{{Cromartie}
  et~al.}{2020}]{Cromartie20}
{Cromartie} H.~T.,  et~al., 2020, \mn@doi [Nature Astronomy]
  {10.1038/s41550-019-0880-2}, \href
  {https://ui.adsabs.harvard.edu/abs/2020NatAs...4...72C} {4, 72}

\bibitem[\protect\citeauthoryear{{De}, {Finstad}, {Lattimer}, {Brown}, {Berger}
   \& {Biwer}}{{De} et~al.}{2018}]{De18}
{De} S.,  {Finstad} D.,  {Lattimer} J.~M.,  {Brown} D.~A.,  {Berger} E.,
  {Biwer} C.~M.,  2018, \mn@doi [Physical Review Letters]
  {10.1103/PhysRevLett.121.091102}, \href
  {http://adsabs.harvard.edu/abs/2018PhRvL.121i1102D} {121, 091102}

\bibitem[\protect\citeauthoryear{{Dietrich}, {Samajdar}, {Khan},
  {Johnson-McDaniel}, {Dudi}  \& {Tichy}}{{Dietrich} et~al.}{2019}]{Dietrich19}
{Dietrich} T.,  {Samajdar} A.,  {Khan} S.,  {Johnson-McDaniel} N.~K.,  {Dudi}
  R.,   {Tichy} W.,  2019, \mn@doi [\prd] {10.1103/PhysRevD.100.044003}, \href
  {https://ui.adsabs.harvard.edu/abs/2019PhRvD.100d4003D} {100, 044003}

\bibitem[\protect\citeauthoryear{{Drout} et~al.,}{{Drout}
  et~al.}{2017}]{Drout17}
{Drout} M.~R.,  et~al., 2017, \mn@doi [Science] {10.1126/science.aaq0049},
  \href {http://adsabs.harvard.edu/abs/2017Sci...358.1570D} {358, 1570}

\bibitem[\protect\citeauthoryear{{Ertl}, {Woosley}, {Sukhbold}  \&
  {Janka}}{{Ertl} et~al.}{2019}]{Ertl19}
{Ertl} T.,  {Woosley} S.~E.,  {Sukhbold} T.,   {Janka} H.~T.,  2019, arXiv
  e-prints, \href {https://ui.adsabs.harvard.edu/abs/2019arXiv191001641E} {p.
  arXiv:1910.01641}

\bibitem[\protect\citeauthoryear{{Evans} et~al.,}{{Evans}
  et~al.}{2017}]{Evans17}
{Evans} P.~A.,  et~al., 2017, \mn@doi [Science] {10.1126/science.aap9580},
  \href {http://adsabs.harvard.edu/abs/2017Sci...358.1565E} {358, 1565}

\bibitem[\protect\citeauthoryear{{Farr}, {Sravan}, {Cantrell}, {Kreidberg},
  {Bailyn}, {Mandel}  \& {Kalogera}}{{Farr} et~al.}{2011}]{Farr11}
{Farr} W.~M.,  {Sravan} N.,  {Cantrell} A.,  {Kreidberg} L.,  {Bailyn} C.~D.,
  {Mandel} I.,   {Kalogera} V.,  2011, \mn@doi [\apj]
  {10.1088/0004-637X/741/2/103}, \href
  {https://ui.adsabs.harvard.edu/abs/2011ApJ...741..103F} {741, 103}

\bibitem[\protect\citeauthoryear{{Farrow}, {Zhu}  \& {Thrane}}{{Farrow}
  et~al.}{2019}]{Farrow19}
{Farrow} N.,  {Zhu} X.-J.,   {Thrane} E.,  2019, \mn@doi [\apj]
  {10.3847/1538-4357/ab12e3}, \href
  {https://ui.adsabs.harvard.edu/abs/2019ApJ...876...18F} {876, 18}

\bibitem[\protect\citeauthoryear{{Ferdman} \& {PALFA Collaboration}}{{Ferdman}
  \& {PALFA Collaboration}}{2018}]{Ferdman18}
{Ferdman} R.~D.,  {PALFA Collaboration} 2018, in {Weltevrede} P.,  {Perera}
  B.~B.~P.,  {Preston} L.~L.,   {Sanidas} S.,  eds,  IAU Symposium Vol. 337,
  Pulsar Astrophysics the Next Fifty Years. pp 146--149,
  \mn@doi{10.1017/S1743921317009139}

\bibitem[\protect\citeauthoryear{{Ferdman} et~al.,}{{Ferdman}
  et~al.}{2014}]{Ferdman14}
{Ferdman} R.~D.,  et~al., 2014, \mn@doi [\mnras] {10.1093/mnras/stu1223}, \href
  {https://ui.adsabs.harvard.edu/abs/2014MNRAS.443.2183F} {443, 2183}

\bibitem[\protect\citeauthoryear{{Fern{\'a}ndez} \& {Metzger}}{{Fern{\'a}ndez}
  \& {Metzger}}{2013}]{Fernandez13}
{Fern{\'a}ndez} R.,  {Metzger} B.~D.,  2013, \mn@doi [\mnras]
  {10.1093/mnras/stt1312}, \href
  {https://ui.adsabs.harvard.edu/abs/2013MNRAS.435..502F} {435, 502}

\bibitem[\protect\citeauthoryear{{Freiburghaus}, {Rosswog}  \&
  {Thielemann}}{{Freiburghaus} et~al.}{1999}]{Freiburghaus99}
{Freiburghaus} C.,  {Rosswog} S.,   {Thielemann} F.-K.,  1999, \mn@doi [\apjl]
  {10.1086/312343}, \href {http://adsabs.harvard.edu/abs/1999ApJ...525L.121F}
  {525, L121}

\bibitem[\protect\citeauthoryear{{Goldstein} et~al.,}{{Goldstein}
  et~al.}{2017}]{Goldstein17}
{Goldstein} A.,  et~al., 2017, \mn@doi [\apjl] {10.3847/2041-8213/aa8f41},
  \href {https://ui.adsabs.harvard.edu/abs/2017ApJ...848L..14G} {848, L14}

\bibitem[\protect\citeauthoryear{{Gompertz} et~al.,}{{Gompertz}
  et~al.}{2018}]{Gompertz18}
{Gompertz} B.~P.,  et~al., 2018, \mn@doi [\apj] {10.3847/1538-4357/aac206},
  \href {http://adsabs.harvard.edu/abs/2018ApJ...860...62G} {860, 62}

\bibitem[\protect\citeauthoryear{{Grevesse}, {Asplund}  \& {Sauval}}{{Grevesse}
  et~al.}{2007}]{Grevesse07}
{Grevesse} N.,  {Asplund} M.,   {Sauval} A.~J.,  2007, \mn@doi [\ssr]
  {10.1007/s11214-007-9173-7}, \href
  {https://ui.adsabs.harvard.edu/abs/2007SSRv..130..105G} {130, 105}

\bibitem[\protect\citeauthoryear{{Guidorzi} et~al.,}{{Guidorzi}
  et~al.}{2017}]{Guidorzi17}
{Guidorzi} C.,  et~al., 2017, \mn@doi [\apjl] {10.3847/2041-8213/aaa009}, \href
  {http://adsabs.harvard.edu/abs/2017ApJ...851L..36G} {851, L36}

\bibitem[\protect\citeauthoryear{{Han}, {Tang}, {Hu}, {Li}, {Jiang}, {Jin},
  {Fan}  \& {Wei}}{{Han} et~al.}{2020}]{Han20}
{Han} M.-Z.,  {Tang} S.-P.,  {Hu} Y.-M.,  {Li} Y.-J.,  {Jiang} J.-L.,  {Jin}
  Z.-P.,  {Fan} Y.-Z.,   {Wei} D.-M.,  2020, arXiv e-prints, \href
  {https://ui.adsabs.harvard.edu/abs/2020arXiv200107882H} {p. arXiv:2001.07882}

\bibitem[\protect\citeauthoryear{{Hill} et~al.,}{{Hill} et~al.}{2002}]{Hill02}
{Hill} V.,  et~al., 2002, \mn@doi [\aap] {10.1051/0004-6361:20020434}, \href
  {https://ui.adsabs.harvard.edu/abs/2002A&A...387..560H} {387, 560}

\bibitem[\protect\citeauthoryear{{Holmbeck}, {Sprouse}, {Mumpower}, {Vassh},
  {Surman}, {Beers}  \& {Kawano}}{{Holmbeck} et~al.}{2019}]{Holmbeck19}
{Holmbeck} E.~M.,  {Sprouse} T.~M.,  {Mumpower} M.~R.,  {Vassh} N.,  {Surman}
  R.,  {Beers} T.~C.,   {Kawano} T.,  2019, \mn@doi [\apj]
  {10.3847/1538-4357/aaefef}, \href
  {https://ui.adsabs.harvard.edu/abs/2019ApJ...870...23H} {870, 23}

\bibitem[\protect\citeauthoryear{{Hosseinzadeh} et~al.,}{{Hosseinzadeh}
  et~al.}{2019}]{Hosseinzadeh19}
{Hosseinzadeh} G.,  et~al., 2019, \mn@doi [\apjl] {10.3847/2041-8213/ab271c},
  \href {https://ui.adsabs.harvard.edu/abs/2019ApJ...880L...4H} {880, L4}

\bibitem[\protect\citeauthoryear{{Kafle}, {Sharma}, {Lewis}  \&
  {Bland-Hawthorn}}{{Kafle} et~al.}{2014}]{Kafle14}
{Kafle} P.~R.,  {Sharma} S.,  {Lewis} G.~F.,   {Bland-Hawthorn} J.,  2014,
  \mn@doi [\apj] {10.1088/0004-637X/794/1/59}, \href
  {https://ui.adsabs.harvard.edu/abs/2014ApJ...794...59K} {794, 59}

\bibitem[\protect\citeauthoryear{{Kasen}, {Metzger}, {Barnes}, {Quataert}  \&
  {Ramirez-Ruiz}}{{Kasen} et~al.}{2017}]{Kasen17}
{Kasen} D.,  {Metzger} B.,  {Barnes} J.,  {Quataert} E.,   {Ramirez-Ruiz} E.,
  2017, \mn@doi [\nat] {10.1038/nature24453}, \href
  {http://adsabs.harvard.edu/abs/2017Natur.551...80K} {551, 80}

\bibitem[\protect\citeauthoryear{{Kasliwal} et~al.,}{{Kasliwal}
  et~al.}{2017}]{Kasliwal17}
{Kasliwal} M.~M.,  et~al., 2017, \mn@doi [Science] {10.1126/science.aap9455},
  \href {http://adsabs.harvard.edu/abs/2017Sci...358.1559K} {358, 1559}

\bibitem[\protect\citeauthoryear{{Kilpatrick} et~al.,}{{Kilpatrick}
  et~al.}{2017}]{Kilpatrick17:gw}
{Kilpatrick} C.~D.,  et~al., 2017, \mn@doi [Science] {10.1126/science.aaq0073},
  \href {https://ui.adsabs.harvard.edu/abs/2017Sci...358.1583K} {358, 1583}

\bibitem[\protect\citeauthoryear{{Kitaura}, {Janka}  \&
  {Hillebrandt}}{{Kitaura} et~al.}{2006}]{Kitaura06}
{Kitaura} F.~S.,  {Janka} H.-T.,   {Hillebrandt} W.,  2006, \mn@doi [\aap]
  {10.1051/0004-6361:20054703}, \href
  {http://adsabs.harvard.edu/abs/2006A%26A...450..345K} {450, 345}

\bibitem[\protect\citeauthoryear{{Korobkin}, {Rosswog}, {Arcones}  \&
  {Winteler}}{{Korobkin} et~al.}{2012}]{Korobkin12}
{Korobkin} O.,  {Rosswog} S.,  {Arcones} A.,   {Winteler} C.,  2012, \mn@doi
  [\mnras] {10.1111/j.1365-2966.2012.21859.x}, \href
  {https://ui.adsabs.harvard.edu/abs/2012MNRAS.426.1940K} {426, 1940}

\bibitem[\protect\citeauthoryear{{Lattimer} \& {Schramm}}{{Lattimer} \&
  {Schramm}}{1974}]{Lattimer74}
{Lattimer} J.~M.,  {Schramm} D.~N.,  1974, \mn@doi [\apjl] {10.1086/181612},
  \href {http://adsabs.harvard.edu/abs/1974ApJ...192L.145L} {192, L145}

\bibitem[\protect\citeauthoryear{{Levan} et~al.,}{{Levan}
  et~al.}{2017}]{Levan17}
{Levan} A.~J.,  et~al., 2017, \mn@doi [\apjl] {10.3847/2041-8213/aa905f}, \href
  {http://adsabs.harvard.edu/abs/2017ApJ...848L..28L} {848, L28}

\bibitem[\protect\citeauthoryear{{Ligo Scientific Collaboration} \& {VIRGO
  Collaboration}}{{Ligo Scientific Collaboration} \& {VIRGO
  Collaboration}}{2019a}]{LVC19:0425_1}
{Ligo Scientific Collaboration} {VIRGO Collaboration} 2019a, GRB Coordinates
  Network, \href {https://ui.adsabs.harvard.edu/abs/2019GCN.24168....1L}
  {24168, 1}

\bibitem[\protect\citeauthoryear{{Ligo Scientific Collaboration} \& {VIRGO
  Collaboration}}{{Ligo Scientific Collaboration} \& {VIRGO
  Collaboration}}{2019b}]{LVC19:0425_2}
{Ligo Scientific Collaboration} {VIRGO Collaboration} 2019b, GRB Coordinates
  Network, \href {https://ui.adsabs.harvard.edu/abs/2019GCN.24228....1L}
  {24228, 1}

\bibitem[\protect\citeauthoryear{{Lippuner} \& {Roberts}}{{Lippuner} \&
  {Roberts}}{2015}]{Lippuner15}
{Lippuner} J.,  {Roberts} L.~F.,  2015, \mn@doi [\apj]
  {10.1088/0004-637X/815/2/82}, \href
  {https://ui.adsabs.harvard.edu/abs/2015ApJ...815...82L} {815, 82}

\bibitem[\protect\citeauthoryear{{Lundquist} et~al.,}{{Lundquist}
  et~al.}{2019}]{Lundquist19}
{Lundquist} M.~J.,  et~al., 2019, \mn@doi [\apjl] {10.3847/2041-8213/ab32f2},
  \href {https://ui.adsabs.harvard.edu/abs/2019ApJ...881L..26L} {881, L26}

\bibitem[\protect\citeauthoryear{{Macias} \& {Ramirez-Ruiz}}{{Macias} \&
  {Ramirez-Ruiz}}{2018}]{Macias18}
{Macias} P.,  {Ramirez-Ruiz} E.,  2018, \mn@doi [\apj]
  {10.3847/1538-4357/aac3e0}, \href
  {https://ui.adsabs.harvard.edu/abs/2018ApJ...860...89M} {860, 89}

\bibitem[\protect\citeauthoryear{{Margalit} \& {Metzger}}{{Margalit} \&
  {Metzger}}{2017}]{Margalit17}
{Margalit} B.,  {Metzger} B.~D.,  2017, \mn@doi [\apjl]
  {10.3847/2041-8213/aa991c}, \href
  {http://adsabs.harvard.edu/abs/2017ApJ...850L..19M} {850, L19}

\bibitem[\protect\citeauthoryear{{Martinez} et~al.,}{{Martinez}
  et~al.}{2015}]{Martinez15}
{Martinez} J.~G.,  et~al., 2015, \mn@doi [\apj] {10.1088/0004-637X/812/2/143},
  \href {https://ui.adsabs.harvard.edu/abs/2015ApJ...812..143M} {812, 143}

\bibitem[\protect\citeauthoryear{{McCully} et~al.,}{{McCully}
  et~al.}{2017}]{McCully17}
{McCully} C.,  et~al., 2017, \mn@doi [\apjl] {10.3847/2041-8213/aa9111}, \href
  {http://adsabs.harvard.edu/abs/2017ApJ...848L..32M} {848, L32}

\bibitem[\protect\citeauthoryear{{Metzger} \& {Fern{\'a}ndez}}{{Metzger} \&
  {Fern{\'a}ndez}}{2014}]{Metzger14}
{Metzger} B.~D.,  {Fern{\'a}ndez} R.,  2014, \mn@doi [\mnras]
  {10.1093/mnras/stu802}, \href
  {http://adsabs.harvard.edu/abs/2014MNRAS.441.3444M} {441, 3444}

\bibitem[\protect\citeauthoryear{{Murguia-Berthier}, {Montes}, {Ramirez-Ruiz},
  {De Colle}  \& {Lee}}{{Murguia-Berthier} et~al.}{2014}]{Murguia-Berthier14}
{Murguia-Berthier} A.,  {Montes} G.,  {Ramirez-Ruiz} E.,  {De Colle} F.,
  {Lee} W.~H.,  2014, \mn@doi [\apjl] {10.1088/2041-8205/788/1/L8}, \href
  {http://adsabs.harvard.edu/abs/2014ApJ...788L...8M} {788, L8}

\bibitem[\protect\citeauthoryear{{Murguia-Berthier} et~al.,}{{Murguia-Berthier}
  et~al.}{2017}]{Murguia-Berthier17:sss17a}
{Murguia-Berthier} A.,  et~al., 2017, \mn@doi [\apjl]
  {10.3847/2041-8213/aa91b3}, \href
  {http://adsabs.harvard.edu/abs/2017ApJ...848L..34M} {848, L34}

\bibitem[\protect\citeauthoryear{{Naiman} et~al.,}{{Naiman}
  et~al.}{2018}]{Naiman18}
{Naiman} J.~P.,  et~al., 2018, \mn@doi [\mnras] {10.1093/mnras/sty618}, \href
  {https://ui.adsabs.harvard.edu/abs/2018MNRAS.477.1206N} {477, 1206}

\bibitem[\protect\citeauthoryear{{Oppenheimer} \& {Volkoff}}{{Oppenheimer} \&
  {Volkoff}}{1939}]{Oppenheimer39}
{Oppenheimer} J.~R.,  {Volkoff} G.~M.,  1939, \mn@doi [Physical Review]
  {10.1103/PhysRev.55.374}, \href
  {https://ui.adsabs.harvard.edu/abs/1939PhRv...55..374O} {55, 374}

\bibitem[\protect\citeauthoryear{{{\"O}zel} \& {Freire}}{{{\"O}zel} \&
  {Freire}}{2016}]{Ozel16}
{{\"O}zel} F.,  {Freire} P.,  2016, \mn@doi [\araa]
  {10.1146/annurev-astro-081915-023322}, \href
  {https://ui.adsabs.harvard.edu/abs/2016ARA&A..54..401O} {54, 401}

\bibitem[\protect\citeauthoryear{{{\"O}zel}, {Psaltis}, {Narayan}  \&
  {McClintock}}{{{\"O}zel} et~al.}{2010}]{Ozel10}
{{\"O}zel} F.,  {Psaltis} D.,  {Narayan} R.,   {McClintock} J.~E.,  2010,
  \mn@doi [\apj] {10.1088/0004-637X/725/2/1918}, \href
  {https://ui.adsabs.harvard.edu/abs/2010ApJ...725.1918O} {725, 1918}

\bibitem[\protect\citeauthoryear{{Pan} et~al.,}{{Pan} et~al.}{2017}]{Pan17:gw}
{Pan} Y.-C.,  et~al., 2017, \mn@doi [\apjl] {10.3847/2041-8213/aa9116}, \href
  {http://adsabs.harvard.edu/abs/2017ApJ...848L..30P} {848, L30}

\bibitem[\protect\citeauthoryear{{Piro}, {Giacomazzo}  \& {Perna}}{{Piro}
  et~al.}{2017}]{Piro17}
{Piro} A.~L.,  {Giacomazzo} B.,   {Perna} R.,  2017, \mn@doi [\apjl]
  {10.3847/2041-8213/aa7f2f}, \href
  {https://ui.adsabs.harvard.edu/abs/2017ApJ...844L..19P} {844, L19}

\bibitem[\protect\citeauthoryear{{Podsiadlowski}, {Langer}, {Poelarends},
  {Rappaport}, {Heger}  \& {Pfahl}}{{Podsiadlowski}
  et~al.}{2004}]{Podsiadlowski04}
{Podsiadlowski} P.,  {Langer} N.,  {Poelarends} A.~J.~T.,  {Rappaport} S.,
  {Heger} A.,   {Pfahl} E.,  2004, \mn@doi [\apj] {10.1086/421713}, \href
  {https://ui.adsabs.harvard.edu/abs/2004ApJ...612.1044P} {612, 1044}

\bibitem[\protect\citeauthoryear{{Pozanenko}, {Minaev}, {Grebenev}  \&
  {Chelovekov}}{{Pozanenko} et~al.}{2019}]{Pozanenko19}
{Pozanenko} A.~S.,  {Minaev} P.~Y.,  {Grebenev} S.~A.,   {Chelovekov} I.~V.,
  2019, arXiv e-prints, \href
  {https://ui.adsabs.harvard.edu/abs/2019arXiv191213112P} {p. arXiv:1912.13112}

\bibitem[\protect\citeauthoryear{{Radice}, {Bernuzzi}, {Del Pozzo}, {Roberts}
  \& {Ott}}{{Radice} et~al.}{2017}]{Radice17}
{Radice} D.,  {Bernuzzi} S.,  {Del Pozzo} W.,  {Roberts} L.~F.,   {Ott} C.~D.,
  2017, \mn@doi [\apjl] {10.3847/2041-8213/aa775f}, \href
  {https://ui.adsabs.harvard.edu/abs/2017ApJ...842L..10R} {842, L10}

\bibitem[\protect\citeauthoryear{{Radice}, {Perego}, {Zappa}  \&
  {Bernuzzi}}{{Radice} et~al.}{2018}]{Radice18:eos}
{Radice} D.,  {Perego} A.,  {Zappa} F.,   {Bernuzzi} S.,  2018, \mn@doi [\apjl]
  {10.3847/2041-8213/aaa402}, \href
  {https://ui.adsabs.harvard.edu/abs/2018ApJ...852L..29R} {852, L29}

\bibitem[\protect\citeauthoryear{{Ramirez-Ruiz}, {Andrews}  \&
  {Schr{\o}der}}{{Ramirez-Ruiz} et~al.}{2019}]{Ramirez-Ruiz19}
{Ramirez-Ruiz} E.,  {Andrews} J.~J.,   {Schr{\o}der} S.~L.,  2019, \mn@doi
  [\apjl] {10.3847/2041-8213/ab3f2c}, \href
  {https://ui.adsabs.harvard.edu/abs/2019ApJ...883L...6R} {883, L6}

\bibitem[\protect\citeauthoryear{{Roederer}, {Kratz}, {Frebel}, {Christlieb},
  {Pfeiffer}, {Cowan}  \& {Sneden}}{{Roederer} et~al.}{2009}]{Roederer09}
{Roederer} I.~U.,  {Kratz} K.-L.,  {Frebel} A.,  {Christlieb} N.,  {Pfeiffer}
  B.,  {Cowan} J.~J.,   {Sneden} C.,  2009, \mn@doi [\apj]
  {10.1088/0004-637X/698/2/1963}, \href
  {https://ui.adsabs.harvard.edu/abs/2009ApJ...698.1963R} {698, 1963}

\bibitem[\protect\citeauthoryear{{Romero-Shaw}, {Farrow}, {Stevenson}, {Thrane}
   \& {Zhu}}{{Romero-Shaw} et~al.}{2020}]{Romero-Shaw20}
{Romero-Shaw} I.~M.,  {Farrow} N.,  {Stevenson} S.,  {Thrane} E.,   {Zhu}
  X.-J.,  2020, arXiv e-prints, \href
  {https://ui.adsabs.harvard.edu/abs/2020arXiv200106492R} {p. arXiv:2001.06492}

\bibitem[\protect\citeauthoryear{{Rosswog}}{{Rosswog}}{2005}]{Rosswog05}
{Rosswog} S.,  2005, \mn@doi [\apj] {10.1086/497062}, \href
  {https://ui.adsabs.harvard.edu/abs/2005ApJ...634.1202R} {634, 1202}

\bibitem[\protect\citeauthoryear{{Rosswog}}{{Rosswog}}{2013}]{Rosswog13}
{Rosswog} S.,  2013, \mn@doi [Philosophical Transactions of the Royal Society
  of London Series A] {10.1098/rsta.2012.0272}, \href
  {https://ui.adsabs.harvard.edu/abs/2013RSPTA.37120272R} {371, 20120272}

\bibitem[\protect\citeauthoryear{{Rosswog}, {Sollerman}, {Feindt}, {Goobar},
  {Korobkin}, {Wollaeger}, {Fremling}  \& {Kasliwal}}{{Rosswog}
  et~al.}{2018}]{Rosswog18}
{Rosswog} S.,  {Sollerman} J.,  {Feindt} U.,  {Goobar} A.,  {Korobkin} O.,
  {Wollaeger} R.,  {Fremling} C.,   {Kasliwal} M.~M.,  2018, \mn@doi [\aap]
  {10.1051/0004-6361/201732117}, \href
  {http://adsabs.harvard.edu/abs/2018A%26A...615A.132R} {615, A132}

\bibitem[\protect\citeauthoryear{{Ruiz}, {Shapiro}  \& {Tsokaros}}{{Ruiz}
  et~al.}{2018}]{Ruiz18}
{Ruiz} M.,  {Shapiro} S.~L.,   {Tsokaros} A.,  2018, \mn@doi [\prd]
  {10.1103/PhysRevD.97.021501}, \href
  {http://adsabs.harvard.edu/abs/2018PhRvD..97b1501R} {97, 021501}

\bibitem[\protect\citeauthoryear{{Safarzadeh}, {Ramirez-Ruiz}  \&
  {Berger}}{{Safarzadeh} et~al.}{2020}]{Safarzadeh20}
{Safarzadeh} M.,  {Ramirez-Ruiz} E.,   {Berger} E.,  2020, arXiv e-prints,
  \href {https://ui.adsabs.harvard.edu/abs/2020arXiv200104502S} {p.
  arXiv:2001.04502}

\bibitem[\protect\citeauthoryear{{Shen}, {Cooke}, {Ramirez-Ruiz}, {Madau},
  {Mayer}  \& {Guedes}}{{Shen} et~al.}{2015}]{Shen15}
{Shen} S.,  {Cooke} R.~J.,  {Ramirez-Ruiz} E.,  {Madau} P.,  {Mayer} L.,
  {Guedes} J.,  2015, \mn@doi [\apj] {10.1088/0004-637X/807/2/115}, \href
  {https://ui.adsabs.harvard.edu/abs/2015ApJ...807..115S} {807, 115}

\bibitem[\protect\citeauthoryear{{Shibata}, {Fujibayashi}, {Hotokezaka},
  {Kiuchi}, {Kyutoku}, {Sekiguchi}  \& {Tanaka}}{{Shibata}
  et~al.}{2017}]{Shibata17}
{Shibata} M.,  {Fujibayashi} S.,  {Hotokezaka} K.,  {Kiuchi} K.,  {Kyutoku} K.,
   {Sekiguchi} Y.,   {Tanaka} M.,  2017, \mn@doi [\prd]
  {10.1103/PhysRevD.96.123012}, \href
  {http://adsabs.harvard.edu/abs/2017PhRvD..96l3012S} {96, 123012}

\bibitem[\protect\citeauthoryear{{Shibata}, {Zhou}, {Kiuchi}  \&
  {Fujibayashi}}{{Shibata} et~al.}{2019}]{Shibata19}
{Shibata} M.,  {Zhou} E.,  {Kiuchi} K.,   {Fujibayashi} S.,  2019, \mn@doi
  [\prd] {10.1103/PhysRevD.100.023015}, \href
  {https://ui.adsabs.harvard.edu/abs/2019PhRvD.100b3015S} {100, 023015}

\bibitem[\protect\citeauthoryear{{Siebert} et~al.,}{{Siebert}
  et~al.}{2017}]{Siebert17}
{Siebert} M.~R.,  et~al., 2017, \mn@doi [\apjl] {10.3847/2041-8213/aa905e},
  \href {http://adsabs.harvard.edu/abs/2017ApJ...848L..26S} {848, L26}

\bibitem[\protect\citeauthoryear{{Smartt} et~al.,}{{Smartt}
  et~al.}{2017}]{Smartt17}
{Smartt} S.~J.,  et~al., 2017, \mn@doi [\nat] {10.1038/nature24303}, \href
  {http://adsabs.harvard.edu/abs/2017Natur.551...75S} {551, 75}

\bibitem[\protect\citeauthoryear{{Song}, {Ai}, {Wang}, {Xing}, {Gao}  \&
  {Zhang}}{{Song} et~al.}{2019}]{Song19}
{Song} H.-R.,  {Ai} S.-K.,  {Wang} M.-H.,  {Xing} N.,  {Gao} H.,   {Zhang} B.,
  2019, \mn@doi [\apjl] {10.3847/2041-8213/ab3921}, \href
  {https://ui.adsabs.harvard.edu/abs/2019ApJ...881L..40S} {881, L40}

\bibitem[\protect\citeauthoryear{{Sukhbold}, {Ertl}, {Woosley}, {Brown}  \&
  {Janka}}{{Sukhbold} et~al.}{2016}]{Sukhbold16}
{Sukhbold} T.,  {Ertl} T.,  {Woosley} S.~E.,  {Brown} J.~M.,   {Janka} H.~T.,
  2016, \mn@doi [\apj] {10.3847/0004-637X/821/1/38}, \href
  {https://ui.adsabs.harvard.edu/abs/2016ApJ...821...38S} {821, 38}

\bibitem[\protect\citeauthoryear{{Suwa}, {Yoshida}, {Shibata}, {Umeda}  \&
  {Takahashi}}{{Suwa} et~al.}{2018}]{Suwa18}
{Suwa} Y.,  {Yoshida} T.,  {Shibata} M.,  {Umeda} H.,   {Takahashi} K.,  2018,
  \mn@doi [\mnras] {10.1093/mnras/sty2460}, \href
  {https://ui.adsabs.harvard.edu/abs/2018MNRAS.481.3305S} {481, 3305}

\bibitem[\protect\citeauthoryear{{Tanaka} et~al.,}{{Tanaka}
  et~al.}{2017}]{Tanaka17}
{Tanaka} M.,  et~al., 2017, \mn@doi [\pasj] {10.1093/pasj/psx121}, \href
  {https://ui.adsabs.harvard.edu/abs/2017PASJ...69..102T} {69, 102}

\bibitem[\protect\citeauthoryear{{Tanvir} et~al.,}{{Tanvir}
  et~al.}{2017}]{Tanvir17}
{Tanvir} N.~R.,  et~al., 2017, \mn@doi [\apjl] {10.3847/2041-8213/aa90b6},
  \href {http://adsabs.harvard.edu/abs/2017ApJ...848L..27T} {848, L27}

\bibitem[\protect\citeauthoryear{{Tauris} \& {Janka}}{{Tauris} \&
  {Janka}}{2019}]{Tauris19}
{Tauris} T.~M.,  {Janka} H.-T.,  2019, \mn@doi [\apjl]
  {10.3847/2041-8213/ab5642}, \href
  {https://ui.adsabs.harvard.edu/abs/2019ApJ...886L..20T} {886, L20}

\bibitem[\protect\citeauthoryear{{The LIGO Scientific Collaboration} \& {the
  Virgo Collaboration}}{{The LIGO Scientific Collaboration} \& {the Virgo
  Collaboration}}{2020}]{Abbott20}
{The LIGO Scientific Collaboration} {the Virgo Collaboration} 2020, arXiv
  e-prints, \href {https://ui.adsabs.harvard.edu/abs/2020arXiv200101761T} {p.
  arXiv:2001.01761}

\bibitem[\protect\citeauthoryear{{The LIGO Scientific Collaboration}
  et~al.,}{{The LIGO Scientific Collaboration} et~al.}{2018}]{Abbott19:sample}
{The LIGO Scientific Collaboration} et~al., 2018, arXiv e-prints, \href
  {http://adsabs.harvard.edu/abs/2018arXiv181112907T} {}

\bibitem[\protect\citeauthoryear{{Timmes}, {Woosley}  \& {Weaver}}{{Timmes}
  et~al.}{1996}]{Timmes96}
{Timmes} F.~X.,  {Woosley} S.~E.,   {Weaver} T.~A.,  1996, \mn@doi [\apj]
  {10.1086/176778}, \href
  {https://ui.adsabs.harvard.edu/abs/1996ApJ...457..834T} {457, 834}

\bibitem[\protect\citeauthoryear{{Tohuvavohu} et~al.,}{{Tohuvavohu}
  et~al.}{2019}]{Tohuvavohu19}
{Tohuvavohu} A.,  et~al., 2019, GRB Coordinates Network, \href
  {https://ui.adsabs.harvard.edu/abs/2019GCN.24353....1T} {24353, 1}

\bibitem[\protect\citeauthoryear{{Tolman}}{{Tolman}}{1939}]{Tolman39}
{Tolman} R.~C.,  1939, \mn@doi [Physical Review] {10.1103/PhysRev.55.364},
  \href {https://ui.adsabs.harvard.edu/abs/1939PhRv...55..364T} {55, 364}

\bibitem[\protect\citeauthoryear{{Ugliano}, {Janka}, {Marek}  \&
  {Arcones}}{{Ugliano} et~al.}{2012}]{Ugliano12}
{Ugliano} M.,  {Janka} H.-T.,  {Marek} A.,   {Arcones} A.,  2012, \mn@doi
  [\apj] {10.1088/0004-637X/757/1/69}, \href
  {https://ui.adsabs.harvard.edu/abs/2012ApJ...757...69U} {757, 69}

\bibitem[\protect\citeauthoryear{{Utsumi} et~al.,}{{Utsumi}
  et~al.}{2017}]{Utsumi17}
{Utsumi} Y.,  et~al., 2017, \mn@doi [\pasj] {10.1093/pasj/psx118}, \href
  {https://ui.adsabs.harvard.edu/abs/2017PASJ...69..101U} {69, 101}

\bibitem[\protect\citeauthoryear{{Valenti} et~al.,}{{Valenti}
  et~al.}{2017}]{Valenti17}
{Valenti} S.,  et~al., 2017, \mn@doi [\apjl] {10.3847/2041-8213/aa8edf}, \href
  {http://adsabs.harvard.edu/abs/2017ApJ...848L..24V} {848, L24}

\bibitem[\protect\citeauthoryear{{Villar} et~al.,}{{Villar}
  et~al.}{2017}]{Villar17}
{Villar} V.~A.,  et~al., 2017, \mn@doi [\apjl] {10.3847/2041-8213/aa9c84},
  \href {http://adsabs.harvard.edu/abs/2017ApJ...851L..21V} {851, L21}

\bibitem[\protect\citeauthoryear{{Wanajo}, {Sekiguchi}, {Nishimura}, {Kiuchi},
  {Kyutoku}  \& {Shibata}}{{Wanajo} et~al.}{2014}]{Wanajo14}
{Wanajo} S.,  {Sekiguchi} Y.,  {Nishimura} N.,  {Kiuchi} K.,  {Kyutoku} K.,
  {Shibata} M.,  2014, \mn@doi [\apjl] {10.1088/2041-8205/789/2/L39}, \href
  {https://ui.adsabs.harvard.edu/abs/2014ApJ...789L..39W} {789, L39}

\bibitem[\protect\citeauthoryear{{van de Voort}, {Quataert}, {Hopkins},
  {Kere{\v{s}}}  \& {Faucher-Gigu{\`e}re}}{{van de Voort}
  et~al.}{2015}]{VandeVoort15}
{van de Voort} F.,  {Quataert} E.,  {Hopkins} P.~F.,  {Kere{\v{s}}} D.,
  {Faucher-Gigu{\`e}re} C.-A.,  2015, \mn@doi [\mnras] {10.1093/mnras/stu2404},
  \href {https://ui.adsabs.harvard.edu/abs/2015MNRAS.447..140V} {447, 140}

\bibitem[\protect\citeauthoryear{{van de Voort}, {Pakmor}, {Grand },
  {Springel}, {G{\'o}mez}  \& {Marinacci}}{{van de Voort}
  et~al.}{2019}]{VandeVoort19}
{van de Voort} F.,  {Pakmor} R.,  {Grand } R. J.~J.,  {Springel} V.,
  {G{\'o}mez} F.~A.,   {Marinacci} F.,  2019, arXiv e-prints, \href
  {https://ui.adsabs.harvard.edu/abs/2019arXiv190701557V} {p. arXiv:1907.01557}

\makeatother
\end{thebibliography}

\onecolumn

\begin{deluxetable}{llllll}
\renewcommand{\arraystretch}{1.4}
\tabletypesize{\footnotesize}
\tablewidth{0pt}
\tablecaption{Source properties for GW190425\label{t:prop}}
\tablehead{
\colhead{Parameter} &
\colhead{LVC high-spin prior} &
\colhead{BNS} &
\colhead{BNS + $m_{2} \approx M_{\rm Ch}$} &
\colhead{BNS + $q \approx 1$} &
\colhead{NSBH}}

\startdata

Primary mass $m_{1}$ (M$_{\sun}$)                  & \lvcmone  & \bnsmone  & \bnsmchmone  & \bnsqmone  & \nsbhmone \\
Secondary mass $m_{2}$ (M$_{\sun}$)                & \lvcmtwo  & \bnsmtwo  & \bnsmchmtwo  & \bnsqmtwo  & \nsbhmtwo \\
Chirp mass $\mathcal{M}$ (M$_{\sun}$)              & \lvcchirp & \bnschirp & \bnsmchchirp & \bnsqchirp & \nsbhchirp\\
Mass ratio $q = m_{2}/m_{1}$                       & \lvcq     & \bnsq     & \bnsmchq     & \bnsqq     & \nsbhq    \\
Total mass $m_{\rm tot}$ (M$_{\sun}$)              & \lvcmtot  & \bnsmtot  & \bnsmchmtot  & \bnsqmtot  & \nsbhmtot \\
Effective inspiral spin parameter $\chi_{\rm eff}$ & \lvcchi   & \bnschi   & \bnsmchchi   & \bnsqchi   & \nsbhchi  \\
Luminosity distance $D_{L}$ (Mpc)                  & \lvcdl    & \bnsdl    & \bnsmchdl    & \bnsqdl    & \nsbhdl    

\enddata

\end{deluxetable}


\end{document}